\documentclass[
reprint,
twocolumn,
aps,
prl,
citeautoscript,
showpacs,
showkeys,
groupedaddress,
superscriptaddress,
twocolumn
amsmath,
amssymb,
]{revtex4-2}

\usepackage{graphicx}
\usepackage{dcolumn}
\usepackage{bm}
\usepackage[hidelinks]{hyperref}
\hypersetup{
    colorlinks,
    linkcolor={blue!50!black},
    citecolor={blue!50!black},
    urlcolor={blue!80!black}
}

\usepackage{siunitx}
\usepackage{xcolor}
\usepackage{braket}

\usepackage{mathtools}
\newcommand{\corrtime}[0]{\ensuremath{\tau_\mathrm{c}}}
\newcommand{\decorrrate}[0]{\ensuremath{1 / \corrtime{}}}
\newcommand{\disorderstrength}[0]{\ensuremath{\langle V \rangle}}

\newcommand{\boltzmann}{k_\mathrm{B}}
\newcommand{\avg}[1]{\langle #1 \rangle}

\begin{document}

\title{Stabilizing an ultracold Fermi gas against\\
Fermi acceleration to superdiffusion through localization}

\author{S.~Barbosa}
\affiliation{Department of Physics and Research Center OPTIMAS, RPTU Kaiserslautern-Landau, 67663 Kaiserslautern, Germany}

\author{M.~Kiefer-Emmanouilidis}
\affiliation{Department of Physics and Research Center OPTIMAS, RPTU Kaiserslautern-Landau, 67663 Kaiserslautern, Germany}
\affiliation{Department of Computer Science and Research Initiative QC-AI, RPTU Kaiserslautern-Landau, 67663 Kaiserslautern, Germany}
\affiliation{Embedded Intelligence, German Research Centre for Artificial Intelligence, 67663 Kaiserslautern, Germany}

\author{F.~Lang}
\affiliation{Department of Physics and Research Center OPTIMAS, RPTU Kaiserslautern-Landau, 67663 Kaiserslautern, Germany}

\author{J.~Koch}
\affiliation{Department of Physics and Research Center OPTIMAS, RPTU Kaiserslautern-Landau, 67663 Kaiserslautern, Germany}

\author{A.~Widera}
\email[]{widera@rptu.de}
\affiliation{Department of Physics and Research Center OPTIMAS, RPTU Kaiserslautern-Landau, 67663 Kaiserslautern, Germany}

\date{\today}

\begin{abstract}
Anderson localization, \textit{i.e.}, destructive quantum interference of multiple-scattering paths, halts transport entirely. 
Contrarily, time-dependent random forces expedite transport via Fermi acceleration, proposed as a mechanism for high-energy cosmic rays. 
Their competition creates interesting dynamics, but experimental observations are scarce. 
Here, we experimentally study the expansion of an ultracold Fermi gas inside time-dependent disorder and observe distinct regimes from sub- to superdiffusion. 
Unexpectedly, quantum interference counteracts acceleration in strong disorder. 
Our system enables the investigation of Fermi acceleration in the quantum-transport regime. 
\end{abstract}
\maketitle

Brownian diffusion, its microscopic understanding, and its application to macroscopic problems enabled the emergence and development of modern science~\cite{einstein_investigations_1956dover, duplantier_brownian_2006}. 
Commonly, the diffusive motion of a particle inside a medium is characterized in $d$ dimensions by its position variance $\sigma^2(t) - \sigma^2(0) = 2 d D t$ increasing linearly with time $t$ and diffusion coefficient $D$. 
While successful for the description of diffusion in many fields, some highly interesting phenomena are reflected by deviations from this behavior and studied extensively in a plethora of systems~\cite{metzler_random_2000, munoz-gil_objective_2021}, including cosmic rays~\cite{uchaikin_fractional_2017}, the foraging behavior of animals~\cite{viswanathan_levy_1996, edwards_revisiting_2007}, fluctuations of the stock market~\cite{plerou_economic_2000}, transport in turbulent plasma~\cite{balescu_anomalous_1995}, and the movement of molecules inside a cell~\cite{di_pierro_anomalous_2018}. 
Different regimes of this so-called anomalous diffusion can be characterized by the value of the diffusion exponent $\alpha$ in a generalized power law~\cite{metzler_random_2000, munoz-gil_objective_2021}
\begin{equation}
\label{eq:anomalous_diffusion}
    \sigma^2(t) - \sigma^2(0) = 2 d D_\alpha t^\alpha,
\end{equation}
where $D_\alpha$ is a generalized diffusion coefficient and the system exhibits subdiffusion (superdiffusion) for $\alpha < 1$ ($\alpha > 1$). 

Perhaps the most extreme form of subdiffusion is the perfect absence of diffusion ($\alpha = 0$), which occurs when quantum particles inside a disordered medium undergo Anderson localization~\cite{anderson_absence_1958, AndersonLocalization}. 
Here,  multiple scattering of a particle's wave function from the disordered environment leads to destructive interference everywhere except for the particle's initial position, resulting in a complete halt of transport. 
Interference-induced localization has been observed in classical waves such as ultrasound~\cite{Weaver1990, Hu2008}, microwaves~\cite{Dalichaouch1991} and light~\cite{Wiersma1997, Scheffold1999, schwartz_transport_2007, Mafi2021} as well as in quantum matter using ultracold atoms~\cite{billy_direct_2008, roati_anderson_2008, jendrzejewski_three-dimensional_2012, kondov_three-dimensional_2011}. 
Further, in three dimensions, a transition from the diffusive to the localized regime occurs at a threshold energy called mobility edge~\cite{AndersonLocalization}. Particles are expected to undergo subdiffusion near that transition point until they become fully localized~\cite{shapiro_cold_2012}. 
Going beyond static disorder, the impact of spatio-temporal noise on localized systems and its transition to delocalization has been investigated intensively~\cite{evensky_localization_1990, lorenzo_remnants_2018}. 
In fact, there has been an interest in the destruction of Anderson localization by temporal variation of an underlying disorder potential in the last decade, particularly in optical systems~\cite{levi_hyper-transport_2012}. 
Remarkably, random time-varying environments or fluctuating force fields have been identified as a major driving force for particle acceleration in outer space, explaining the existence of high-energy cosmic rays~\cite{fermi_origin_1949, sturrock_model_1966, ostrowski_diffusion_1997, mertsch_fermi_2011}. 
This fundamental mechanism, which has come to be known as Fermi acceleration, dates back to Fermi~\cite{fermi_origin_1949} and was later studied in detail for classical and quantum particles scattered in time-varying potentials~\cite{golubovic_classical_1991, rosenbluth_comment_1992, aguer_classical_2010, volpe_brownian_2014}. 
Microscopically, a particle scattered from a co-propagating potential maximum will be decelerated, while it will be accelerated for collisions from counter-propagating potential maxima. 
Statistically, the counter-propagating collisions are more probable with increasing particle velocity. 
Thus, particles moving in time-varying potentials experience an increasing accelerating force. 
This mechanism has since been generalized to the classical Fermi-Ulam-accelerator model~\cite{ulam_statistical_1961, lichtenberg_regular_1983}, which was later expanded to include quantum dynamics~\cite{jose_study_1986, seba_quantum_1990}. 
Moreover, diffusion inside random disorder is expected to exhibit universal behavior with $\sigma^2 \sim t^2$ and $\avg{v^2} \sim t^{2/5}$~\cite{golubovic_classical_1991, rosenbluth_comment_1992, bouchet_minimal_2004, aguer_classical_2010} if $d > 1$, where $\avg{v^2}$ is the variance of velocity. 

\begin{figure*}
    \centering
    \includegraphics[width=178mm]{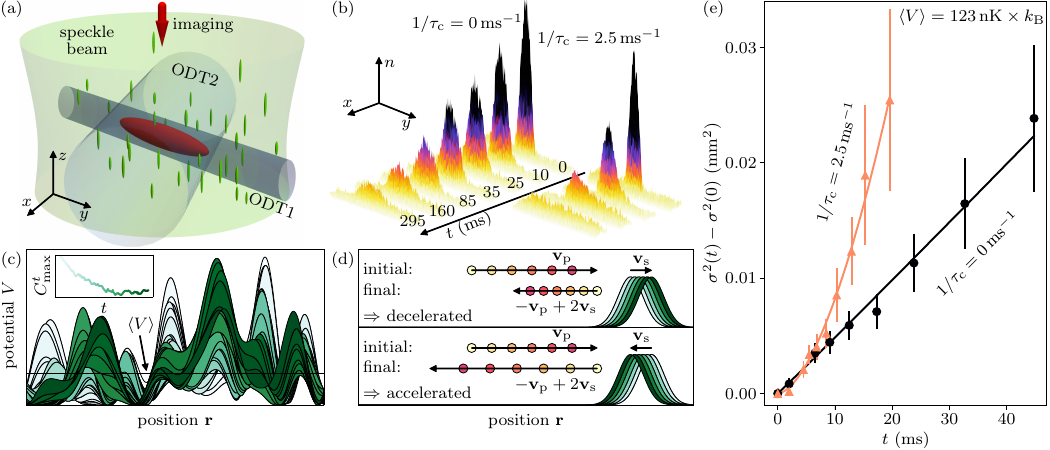}
    \caption{
    (a)~Sketch of atom cloud (red ellipsoid), optical dipole traps (ODT1, ODT2), speckle laser beam (green), and anisotropic speckle grains (small green ellipsoids). Absorption imaging is performed along the $-z$ direction (red arrow). 
    (b)~Surface plot of recorded density distribution for different expansion duration $t$ for $\decorrrate = \SI{0}{\per\milli\second}$ (left) and $\decorrrate = \SI{2.5}{\per\milli\second}$ (right). The length along the $y$ axis is \SI{1}{\milli\meter}. For each setting of $t$ and \decorrrate{}, 50 repetitions are averaged. For increased image clarity, the illustration was smoothed by a Gaussian filter with a standard deviation of one pixel only for this figure. 
    (c)~Visualization of the disorder's time evolution from a 1D numerical simulation. Lines are snapshots along the evolution, later times correspond to darker green. Both peak heights and positions vary with different rates. The black horizontal line marks the average disorder potential \disorderstrength{}, which is constant in $t$. 
    Inset is the cross-correlation peak height $C^t_\mathrm{max}$ in arbitrary units over time $t$ (same color scale)~\cite{nagler_ultracold_2022, hiebel_characterizing_2024}. 
    (d)~Illustration of Fermi acceleration: Read-end collisions (top) effectively decelerate, while head-on collisions accelerate the particle. Colors indicate time as before. 
    (e)~Evolution of measured cloud variance over time from the two data sets shown in (b). Power-law fits (lines) to the data (points) yield diffusion exponents of $\alpha = 1.02\, \pm \,0.04$ for $\decorrrate = \SI{0}{\per\milli\second}$ (circles) and $\alpha = 1.70\, \pm \,0.08$ for $\decorrrate = \SI{2.5}{\per\milli\second}$. 
    Error bars indicate $1\sigma$ statistical uncertainty. 
    }
    \label{fig1}
\end{figure*}

Here, we investigate the expansion of an ultracold non-interacting Fermi gas in a time-varying disorder potential to study the influence of the competing contributions of localization due to matter-wave interference and the acceleration due to stochastic Fermi acceleration on the diffusion of the gas.

Experimentally, we start by producing a degenerate Fermi gas of $^6$Li atoms at a temperature $T \approx \SI{0.1}{} \, T_\mathrm{F} < \SI{100}{\nano K}$, with Fermi temperature $T_\mathrm{F} = E_\mathrm{F} / \boltzmann$, Fermi energy $E_\mathrm{F}$ and Boltzmann constant $\boltzmann$. 
All $N \approx 10^5$ atoms are prepared spin-polarized in the lowest-lying Zeeman substate. 
Our sample is a good approximation of an ideal Fermi gas due to its fermionic nature, as $s$-wave interactions are prohibited entirely due to the Pauli exclusion principle and $p$-wave and higher-order interactions are strongly suppressed at these low temperatures~\cite{shapiro_cold_2012}. 
Initially, the atoms are prepared in a trap created by superposing the main optical dipole trap (ODT1), formed by a focused laser beam propagating along the $y$ axis, with a secondary beam (ODT2), crossing the main beam at an angle in the $x$-$y$ plane, see Fig.~\ref{fig1}(a). 
By extinguishing the ODT2 at time $t=0$, the trap instantly becomes shallow along the $y$ axis while the remaining directions are effectively unchanged. 
Hence, the atoms start to expand along the $y$-direction, see Fig.~\ref{fig1}(b). 
After a variable expansion duration, we extract information about the cloud's extension by performing absorption imaging along the $z$ axis~\cite{ganger_versatile_2018}. 

Simultaneously to switching off the ODT2, at $t = 0$, we quench on a repulsive optical speckle disorder potential $V(\mathbf{r})$, created by \SI{532}{\nano\meter} laser light, see Refs.~\cite{nagler_ultracold_2022, hiebel_characterizing_2024} for details. 
Spatially, it consists of anisotropic grains with typical sizes of $\eta_{x,y}^2 \times \eta_z = (\SI{750}{\nano\meter})^2 \times \SI{10.2}{\micro\meter}$, where $\eta_{x,y}$ and $\eta_z$ are the correlation lengths along the respective directions~\cite{nagler_observing_2022, kuhn_coherent_2007}. 
We characterize the strength of the disorder by its spatial average $\avg{V}$. 
Technically, we can tune the rate \decorrrate{} at which the disorder decorrelates and another speckle realization emerges that shares no resemblance to the original, see Fig.~\ref{fig1}(c). 
While the local details of the disorder potential change significantly with time, its statistical properties, such as correlation length or mean potential, do not. 
Hence, we realize a time-varying stochastic force field for our atom cloud, allowing for stochastic Fermi acceleration (Fig.~\ref{fig1}(d)), 
for details see Refs.~\cite{nagler_ultracold_2022, hiebel_characterizing_2024}. 
In the following, we realize decorrelation rates up to $\decorrrate = \SI{3.5}{\per\milli\second}$ to study the effect on the diffusion of non-interacting atoms in either weak ($\disorderstrength = \SI{123}{\nano K} \times \boltzmann \approx 0.2\, E_F$) or stronger ($\disorderstrength = \SI{401}{\nano K} \times \boltzmann \approx 0.5\, E_F$) disorder, see Supplemental Material. 
We note that, even in the static case, our three-dimensional disorder potential does not allow for any classically bound states~\cite{pilati_dilute_2010, sanchez-palencia_disorder-induced_2008}. 
We further emphasize that the atom cloud never crosses into the regimes of dimensionality lower than $d=3$. 
Nevertheless, we still analyze the diffusive expansion only along one dimension, $d = 1$, as the atoms are prohibited from expanding along the $x$ and $z$ directions.

From the absorption images taken, we extract the cloud width $\sigma$. 
While standard methods such as the fitted Gaussian width or the participation ratio work in principle, we use the so-called inverse participation width (see Supplemental Material and Ref.~\cite{barbosa_what_2023} for details), which is particularly suited to compensate for noise of the imaging process, becoming relevant for long expansion times, when the local density of the cloud decreases. 

\begin{figure}
    \centering
    \includegraphics[width=86mm]{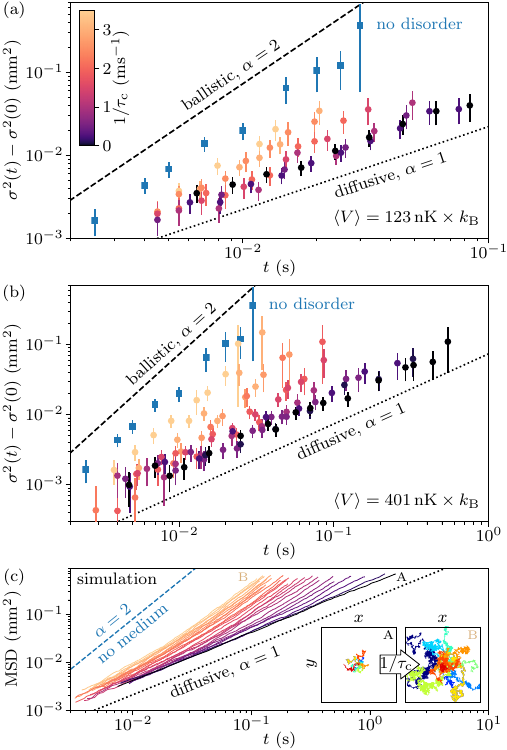}
    \caption{
    Cloud variances from experimental data (points) for different decorrelation rates, increasing from $\decorrrate = \SI{0}{\per\milli\second}$ to $\decorrrate = \SI{3.5}{\per\milli\second}$ (color bar in (a)) for (a)~weak disorder, $\disorderstrength = \SI{123}{\nano K} \times \boltzmann$, and (b)~strong disorder, $\disorderstrength = \SI{401}{\nano K} \times \boltzmann$. Blue squares show variances from disorder-free expansion. Error bars indicate $1\sigma$ statistical uncertainty. The black dotted (dashed) line indicates the exponent $\alpha$ for normal (ballistic) diffusion. 
    (c)~Particle-averaged mean squared displacement (MSD) from a single simulation run. We simulated 25 different velocity scales of the medium, ranging from the static case (A) to the experimentally maximum-achievable dynamics (B), in the same color scale as for the experimental data. The dashed blue line shows free expansion without a medium. 
    Inset: twelve examples of simulated trajectories up to $t = \SI{100}{\milli\second}$ for the static (A,~left) and maximally dynamic case (B,~right) from which we calculate MSD. Both boxes are of size $\SI{1}{\milli\meter} \times \SI{1}{\milli\meter}$. 
    }
    \label{fig2}
\end{figure}

In Fig.~\ref{fig2}, the cloud variances $\sigma^2$ over time are shown in a double-logarithmic plot for (a) weak and (b) strong disorder, where the exponent $\alpha$ of a power-law Eq.~\eqref{eq:anomalous_diffusion} corresponds to the slope of a straight line. 
The data from the expansion inside static disorder, $\decorrrate = \SI{0}{\per\milli\second}$, is shown as black points. 
It exhibits the lowest exponent for both disorder strengths, allowing for the longest observation times. 
The observation time is technically limited by atom losses or the finite size of the camera detection area and the envelope of the disorder speckle pattern. 
Nevertheless, we observe various transient transport regimes. 
When we increase the decorrelation rate toward its maximum value of $\decorrrate = \SI{3.5}{\per\milli\second}$, we recover a strongly increased slope and, therefore, exponent for both disorder strengths. 
In fact, the slope even takes on the same value as in the disorder-free expansion, \textit{i.e.}, ballistic transport. 
This is conform with the predicted transport universality of $\sigma^2 \sim t^2$ for time-dependent force fields~\cite{golubovic_classical_1991, rosenbluth_comment_1992, bouchet_minimal_2004, aguer_classical_2010}. 

To compare the experimental data to expectations derived from Fermi acceleration, we perform a Markov-chain Monte-Carlo simulation employing the minimal stochastic model for Fermi acceleration of Ref.~\cite{bouchet_minimal_2004}, see Supplemental Material for details. The resulting trajectories are used to compute a mean-squared displacement of the simulated particle, see Fig.~\ref{fig2}(c). 
The simulation yields essentially the same accelerating behavior as seen in the experimental data. 

A quantitative analysis of the change of dynamical regimes seen in Fig.~\ref{fig2} is done by extracting the diffusion exponent $\alpha$ and diffusion coefficient $D_\alpha$ from the different series of each $\disorderstrength$ and $\decorrrate$. 
The exponent $\alpha$ is obtained as the slope from linear regression of the logarithm of both variance and time. 
With that, we calculate the anomalous diffusion coefficient as 
$D_\alpha = \left\langle \left(\sigma_i^2(t) - \sigma_i^2(0) \right) / 2 t^\alpha \right\rangle$,
where the set of values that are averaged is constant in time. 
Note that $D_\alpha$ has the unit $\mathrm{m}^2 \, \mathrm{s}^{-\alpha}$~\cite{metzler_random_2000}. 

\begin{figure*}
    \centering
    \includegraphics[width=178mm]{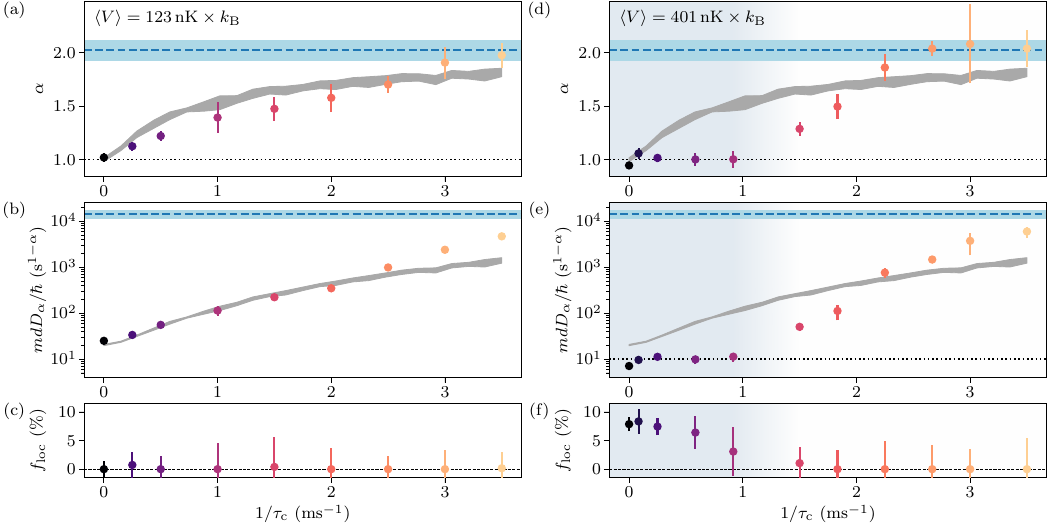}
    \caption{
    Diffusion properties in (a-c)~weak disorder, $\disorderstrength = \SI{123}{\nano K} \times \boltzmann$, and (d-f)~strong disorder, $\disorderstrength = \SI{401}{\nano K} \times \boltzmann$.
    (a,~d)~Diffusion exponents $\alpha$ as a function of decorrelation rate $1/\tau_c$ from experimental measurements (dots) and simulation data of twelve trajectories (gray area). The blue dashed line indicates the disorder-free measurement, and the blue area around it shows its error. Error bars and areas indicate $1\sigma$ statistical uncertainty. 
    (b,~e)~Normalized diffusion coefficient $m d D_\alpha / \hbar$ with the same colors and line types as in panel (a). For the experimental data (simulation), $d = 1$ ($d = 2$), see Supplemental Material for details. Experimental errors are calculated as the standard deviation of values used for averaging. 
    (c,~f)~Localized fraction $f_\mathrm{loc}$. For weak disorder, (c), it is consistent with zero for all decorrelation rates. For strong disorder, (f), we observe a significant $f_\mathrm{loc} > 0$, which vanishes at $\decorrrate \approx \SI{1}{\per\milli\second}$, coinciding with the transition from a constant to a significant increase in both the diffusion exponent and coefficient. The background shading highlights this quantum-transport region in (d-f). 
    }
    \label{fig3}
\end{figure*}

For the expansion in weak disorder, we see a direct and monotonous increase of both $\alpha$ and $D_\alpha$ with \decorrrate{}, see Fig.~\ref{fig3}(a, b). 
Focusing on the exponent, we can tune through the entire range of superdiffusion with exponents between $\alpha = 1$ for static, weak disorder and $\alpha = 2$ by choice of \decorrrate{}. 
The simulation predicts this tunability well, even agreeing quantitatively with the experimental data for a wide range of decorrelation rates. 

Turning to the results for the expansion in strong disorder, we find a strikingly different behavior.
For static disorder, $\decorrrate = \SI{0}{\per\milli\second}$, we find the system to be slightly but statistically significantly in the subdiffusive regime with $\alpha = 0.94\, \pm \,0.03$, see Fig.~\ref{fig3}(d). 
As reported by Ref.~\cite{shapiro_cold_2012}, subdiffusion is expected to occur near the mobility edge until the wave packet has expanded into its fully localized state. 
In our previous work, \textit{i.e.} Ref.~\cite{barbosa_what_2023}, we performed a thorough investigation of the expansion in static disorder of various strengths. 
A widely used estimate for Anderson localization is the Ioffe-Regel criterion, which can be expressed for $^6$Li as $T < \SI{160}{\nano K}$ for the geometric mean of the correlation lengths $\overline{\eta}$ of our speckle disorder ($T < \SI{900}{\nano K}$ for $\eta_{x,y}$ relevant for the expansion direction)~\cite{kondov_three-dimensional_2011}. 
As the gas temperature is $T < \SI{100}{\nano K}$, our system fulfills the Ioffe-Regel criterion. 
An alternative criterion regards the critical momentum $k_\mathrm{AL}$ below which Anderson localization is expected to occur~\cite{beilin_diffusion_2010}. 
Specifically for our case $\disorderstrength / E_\mathrm{c} > 1$, with correlation energy $E_\mathrm{c} = \hbar^2 / (m \eta^2)$, we use the estimation $k_\mathrm{AL} \approx (\disorderstrength / E_\mathrm{c})^{2/5} / \eta$, where $\hbar$ is the reduced Planck constant and $m$ is the atom mass. 
By comparison with the Fermi momentum $k_\mathrm{F}$, the largest momentum present in our degenerate Fermi gas, we get $k_\mathrm{F} \approx 2 \, k_\mathrm{AL}$ for $\overline{\eta}$ (and $k_\mathrm{F} \approx 0.8 \, k_\mathrm{AL}$ for $\eta_{x,y}$). 
In any case, we can expect at least a significant fraction of the low-energy fermions to localize, see also Ref.~\cite{barbosa_what_2023}. 
Therefore, we attribute the onset of subdiffusion for strong disorder as a signature of Anderson localization below the mobility edge, slowing down the expansion. 

This is supported by the experimental observation that the diffusion coefficient is of the order of only a few `quanta of diffusion' $\hbar / m$~\cite{kuhn_coherent_2007, jendrzejewski_three-dimensional_2012, shapiro_cold_2012}, see Fig.~\ref{fig3}(e). 
Furthermore, with increasing $\decorrrate{} \lesssim \SI{1}{\per\milli\second}$, we see an initial plateau where neither diffusion quantity, exponent or coefficient, changes (background shading in Fig.~\ref{fig3}(c-d)). 
It clearly illustrates a strong suppression of the Fermi acceleration for sufficiently slow changes of the underlying disorder potential. 
The experimental observation is also in stark contrast to the classical simulation based on Fermi acceleration.
We interpret this observation as the effect of localization due to wavefunction interference for small decorrelation rates. 

To investigate the interplay of localization effects with superdiffusion or its suppression more closely, we infer a measure for the fraction $f_\mathrm{loc}$ of localized atoms as introduced in Ref.~\cite{jendrzejewski_three-dimensional_2012}. 
It quantifies the fraction of particles that would not have diffused away from the initial cloud volume after infinite time, see Supplemental Material and Ref.~\cite{barbosa_what_2023} for more details. 
We find a localized fraction of zero for the expansion in weak disorder even in the static case, as expected and can be seen in Fig.~\ref{fig3}(c). 
By contrast, we find a significant localized fraction $f_\mathrm{loc} = \SI{7.9 \pm 1.2}{\%}$ for the strong static disorder, see Fig.~\ref{fig3}(f). 
As expected, the fraction decays as the decorrelation rate \decorrrate{} increases, but as long as a localized fraction persists, the system shows close-to-normal diffusion. 
However, the normal-diffusion plateau ends as soon as the localized fraction has decayed. 

We interpret this plateau of $\alpha$ and $D_\alpha$ as a consequence of localization effects stabilizing diffusion against the disorder's accelerating dynamics. 
In fact, the energy scale $h / \corrtime \approx \SI{48}{\nano K} \times \boltzmann$ for the observed delocalization rate $\decorrrate \approx \SI{1}{\per\milli\second}$ is of the same order of magnitude as the energy associated with the critical momentum for Anderson localization, being $E_\mathrm{AL} = \hbar^2 k_\mathrm{AL}^2 / 2m \approx \SI{115}{\nano K} \times \boltzmann$ for the strong disorder and $\overline{\eta}$. 
Therefore, the fraction of particles with sufficiently low energy to localize despite the additional energy $E_\mathrm{AL}$ will be increasingly reduced with $\decorrrate$ until too few low-energy particles remain to influence the transport globally. 
Around $\decorrrate \approx \SI{1}{\per\milli\second}$, where $f_\mathrm{loc}$ has vanished, the Fermi acceleration becomes too strong to sustain coherent matter-wave interference and finally drives the system to superdiffusion. 
For the three largest values of \decorrrate{}, we again observe $\alpha = 2$, consistent with the expected universal transport in time-dependent force fields, \textit{i.e.} Fermi acceleration~\cite{golubovic_classical_1991, rosenbluth_comment_1992, bouchet_minimal_2004, aguer_classical_2010}

We have demonstrated a system that exhibits a broad tunability of anomalous diffusion during experimentally achievable timescales, ranging from subdiffusion to superdiffusion, while reaching the regime of ballistic transport. 
Our findings elucidate the intriguing dynamics of matter waves in time-varying random force fields, establishing an experimental platform to investigate Fermi acceleration in quantum systems. 
An interesting prospect will be to achieve decorrelation timescales faster than the inverse Fermi energy. 
Then, we can explore the maximum achievable rate of Fermi acceleration experimentally and discern whether our system tends asymptotically to the universal $\alpha = 2$~\cite{aguer_classical_2010} or if hyper transport as in Ref.~\cite{levi_hyper-transport_2012} is achievable. 
Moreover, it will be interesting in future studies to explore the contributions of particle interactions or superfluidity, as our system is generally capable of creating a strongly interacting Fermi gas along the crossover from a molecular Bose-Einstein condensate (BEC) to a Bardeen-Cooper-Schrieffer (BCS) superfluid~\cite{ganger_versatile_2018, koch_quantum_2023}. 
Finally, this degree of tunable anomalous diffusion might be of value in related applications of wave phenomena, such as atomtronics, electronics, and (electro)chemical settings, where precise control over the transport velocity would be highly desirable. \\

We thank M.~Fleischhauer, C.A.R.~S\'a de Melo, A.~Buchleitner, T.~Enss, D.~Hern\'andez-Rajkov, G.~Roati, and E.~Lutz for discussions as well as B.~Moser, I.~Cardoso Barbosa, and B.~Nagler for carefully reading the manuscript. 
This work was supported by the German Research Foundation (DFG) through the Collaborative Research Center Sonderforschungsbereich SFB/TR185 (Project 277625399). M.K.-E. acknowledges support by the Quantum Initiative Rhineland-Palatinate QUIP Research Initiative Quantum Computing for Artificial Intelligence QC-AI.
J.K. acknowledges support by the Max Planck Graduate Center with the Johannes Gutenberg-Universität Mainz. 

\paragraph{Author contributions.}
S.B. and A.W. conceived the research. S.B., F.L., and J.K. ran the experimental apparatus. S.B. took and analyzed the experimental data. M.K.-E. contributed to the analysis. S.B. wrote and analyzed the simulation. S.B. wrote the initial manuscript draft. All authors contributed to the interpretation of the data, writing of the manuscript and critical feedback. 

\paragraph{Data availability.}
All data of the figures in the manuscript and Supplemental Material are available in a Zenodo repository (Ref.~\cite{zenodo_dynamic}): \href{https://zenodo.org/doi/10.5281/zenodo.10478890}{https://zenodo.org/doi/10.5281/zenodo.10478890}.

\clearpage 

% SUPPLEMENTAL MATERIAL

\onecolumngrid

\begin{center}
    \textbf{\large Supplemental Material for\\
    ``Stabilizing an ultracold Fermi gas against\\
    Fermi acceleration to superdiffusion through localization''}\\[.4cm]
    S.~Barbosa,$^{1}$ M.~Kiefer-Emmanouilidis,$^{1,2,3}$ F.~Lang,$^{1}$ J.~Koch,$^{1}$ and A.~Widera$^{1,*}$\\[.1cm]
    {\small \itshape 
    $^{1}$Department of Physics and Research Center OPTIMAS,\\
    RPTU Kaiserslautern-Landau, 67663 Kaiserslautern, Germany\\
    $^{2}$Department of Computer Science and Research Initiative QC-AI,\\
    RPTU Kaiserslautern-Landau, 67663 Kaiserslautern, Germany\\
    $^{3}$Embedded Intelligence, German Research Center for Artificial Intelligence, 67663 Kaiserslautern, Germany
    }\\[.8cm]
\end{center}

\setcounter{equation}{0}
\setcounter{section}{0}
\setcounter{figure}{0}
\setcounter{page}{1}
\renewcommand{\theequation}{S\arabic{equation}}
\renewcommand{\thesection}{\arabic{section}}
\renewcommand{\thefigure}{S\arabic{figure}}
\renewcommand{\bibnumfmt}[1]{[S#1]}
\renewcommand{\citenumfont}[1]{S#1}
\makeatletter

\twocolumngrid

\section{Experimental details}

Our experimental sequence for the preparation of the spin-polarized gas, the trap configuration, gas temperature, and imaging are based on our previous work Ref.~\cite{barbosa_what_2023S}. 
There, many of the relevant technical details are described comprehensively. 
We employ resonant high-intensity absorption imaging, see Refs.~\cite{reinaudi_strong_2007S, nagler_cloud_2020S}. 
Before the spin polarization, we evaporatively cool a previously laser-cooled sample of $^6$Li atoms in the two lowest Zeeman substates to temperatures around $T = \SI{100}{\nano K}$~\cite{ganger_versatile_2018S}. 
The evaporation takes place at a magnetic field of $B = \SI{763.6}{G}$, on the BEC side of the broad Feshbach resonance, which allows us to tune the interaction strength of the gas such that the cooling is efficient~\cite{grimm_ultracold_2007S, zurn_precise_2013S}. 
Afterward, we adiabatically ramp the magnetic field to $B = \SI{1070}{G}$, deep into the BCS regime, where the fermionic pairs are weakly bound and spatially far apart. 
All atoms in one of the spin states are then removed from the trap by a resonant laser pulse, leaving a spin-polarized sample forming a quasi-pure Fermi sea.
Only about \SI{10}{\%} of the atoms in the lowest-lying state are lost due to resonant scattering, while no measurable amount in the other state remains. 

Since the magnetic field's curvature strongly traps the atoms in the $x$-$y$ plane while being anti-confining along the $z$ axis~\cite{nagler_observing_2022S}, we load the atoms into the crossed dipole trap as shown in Fig.~\ref{fig1}, and switch off the magnetic field. 
At that point, we can either extract the cloud's temperature with the method described in Refs.~\cite{hadzibabic_twospecies_2002S, kinast_phdS} or begin the expansion sequence by switching off ODT2 while switching on the disorder potential, both during less then one microsecond using acousto-optical modulators, faster than the timescale of the motion of the atoms. 
We ensured that, for every step of the preparation sequence, no observable excitations are performed and no significant amount of atoms are lost, the only exception being the polarization pulse as mentioned. 

When driving the system with non-zero correlation rates, we observed that atoms were so strongly accelerated that the cloud expanded significantly along the $x$ direction. 
Therefore, we increased the power of the ODT1 laser from \SI{100}{\milli\watt}, as was used for the weak disorder in this work and for the entirety of Ref.~\cite{barbosa_what_2023S}, to \SI{300}{\milli\watt}. 
The resulting trap parameters for the series with weak disorder amount to $(\omega_x, \omega_y, \omega_z) = (365, 1.9, 248) \times 2\pi \, \SI{}{Hz}$ after $t=0$ and, with that, $E_\mathrm{F} \approx \SI{600}{\nano K}\times\boltzmann$ as well as $\disorderstrength / E_\mathrm{F} \approx 0.2$. 
For the strong-disorder measurements, $(\omega_x, \omega_y, \omega_z) = (670, 3.4, 435) \times 2\pi \, \SI{}{Hz}$, from which we calculate $E_\mathrm{F} \approx \SI{850}{\nano K}\times\boltzmann$ as well as $\disorderstrength / E_\mathrm{F} \approx 0.5$. 
Note that, for both series, $\omega_y = 37.8 \times 2 \pi \, \SI{}{Hz}$ for $t<0$. 
We confirmed that the disorder-free expansion is still ballistic. 
For sufficiently slow dynamics, the expansion in strong disorder is very similar for both trap configurations. 
The energy can be dissipated into the unobserved directions only for the larger decorrelation rates, effectively reducing the superdiffusion along $y$. 

As mentioned in the main text, the expansion time $t$ is limited technically. 
While the cloud grows in size during the expansion, we stop the evaluation whenever its size reaches that of the speckle envelope or, for the disorder-free expansion, that of the camera sensor. 
We further limit $t$ when particle losses become significant at more than \SI{10}{\%}. 
The latter is mostly the case for large values of \decorrrate{}. 
Note that we exclusively smoothed the illustration in Fig.~\ref{fig1}(b) with a Gaussian filter and only for an enhanced visibility. 
Except for averaging the 50 images taken for each setting, no further image manipulation was performed for the evaluation.

\section{Determination of the cloud variance}

We use the inverse participation width (IPW), a statistical observable introduced in our previous work Ref.~\cite{barbosa_what_2023S}, as a measure for the cloud's spatial extension. 
For that, we calculate the histogram of an absorption image taken at time $t$, approximating the particle's position probability density function (PDF), and extract its width $w(t)$ as the full range between the largest and lowest recorded densities. 
As camera noise is present and influences the recorded histogram, we have to take it into account. 
For that, we also extract the width $w_\mathrm{noise}$ of a histogram of a noisy image without any atoms but otherwise identical statistics. 
More precisely, the recorded histogram will be the convolution of the noise-free distribution, a bimodal function for most settings, with the PDF of the camera noise. 
In cases such as ours, $w(t) - w_\mathrm{noise}$ will then be a good approximation for the sample's peak density $n(0, t)$. 
Assuming that the cloud spreads as $n(0, t) \sim N(t) / \sigma$, we get the cloud variance from IPW as
\begin{equation}
\label{eq:sigma_observable}
    \sigma^2(t) \approx \frac{1}{2 \pi} \frac{N^2(t)}{(w(t) - w_\mathrm{noise})^2}.
\end{equation}
Here, we have already inserted the squared prefactor $1/2\pi$ from a Gaussian function to allow a quantitative analysis of the diffusion coefficient. 
We emphasize that the prefactor choice is unimportant for our conclusions because even using a box distribution (which is obviously significantly different from our cloud shape), for example, will yield $1/4$, which is still somewhat comparable to $1/2\pi \approx 0.16$. 
The coefficient can therefore be extracted with an uncertainty of order unity. 
To determine the diffusion exponent quantitatively, only $n(0, t) \sim N(t) / \sigma(t)$ needs to be valid. 
For details about IPW, its regimes of validity, and a comparison with established observables, see Ref.~\cite{barbosa_what_2023S}.

\section{Fermi-acceleration simulation}

As stated in the main text, the simulations are based on the model presented in Ref.~\cite{bouchet_minimal_2004S}. 
We simulate classical non-relativistic point particles colliding elastically with hard-sphere scatterers of infinite mass on a flat two-dimensional plane. 
We choose a 2D system since, in 1D, the mechanism of Fermi acceleration is significantly different due to the lack of scattering angles. 
Single scattering events are effectively the same for dimensions larger than one if the scattering angles are assumed uniformly distributed. 
Since our atom cloud is three-dimensional, we simulate in $d > 1$ and choose $d=2$ as a compromise to save computing resources. 

In the case of frozen scattering centers, normal diffusion is the result. 
However, when these scatterers themselves are moving, the particles undergo Fermi acceleration and expand superdiffusively. 
Since here and in contrast to the experimental setup, we do have access to the trajectories $\mathbf{r}(t)$ of each particle, we directly calculate the mean-squared displacement as
\begin{equation}
\label{eq:msd_trajectories}
    \mathrm{MSD}(t) = \avg{\mathbf{r}^2}(t) - \avg{\mathbf{r}^2}(0),
\end{equation}
where $\avg{\cdot}$ denotes the average over the particles. 
For each simulation run, we set \SI{1000}{} particles, each colliding \SI{25000}{} times with the randomly moving spheres. 
Finally, we run such a series twelve times and average the diffusion exponents and coefficients extracted from each series. 

We choose values as close to the experimental setting as possible for the scatterers. 
As their radius, we use the geometric mean of our disorder's correlation lengths $\overline{\eta}$ and use the average distance of speckle peaks, $3\overline{\eta}$, for their density $\rho = 1 / (3\,\overline{\eta})^2$. 
Since the speckle's spatial intensity is exponentially distributed, we assumed that the same holds for the velocity. 
Therefore, the simulated scatterers' velocity is determined randomly with an exponential probability distribution. 
Note that the choice of their velocity distribution has little to no impact on the result. 
Only the average velocity has a significant influence. 
Therefore, we iterate through 25 different values of their average velocity between zero (see black line A in Fig.~\ref{fig2}(c)) and the maximum value (yellow line B) of $v_\mathrm{sim}^\mathrm{max} = \SI{6.3}{\milli\meter\per\second}$. 
That value is estimated from the velocity scale of our maximally dynamic disorder by comparing the present length and timescales $\overline{\eta} / \tau_\mathrm{c}$ for $\decorrrate = \SI{3.5}{\per\milli\second}$. 
For the initial spatial distribution of the point particles, we choose a Gaussian with $\sigma_{x,y}(0) = \SI{50}{\micro\meter}$. 
For their velocity magnitudes, we distribute values between zero and the Fermi velocity $v_\mathrm{F}$ of our experimental system as they would be for an ideal Fermi gas, while the angles are chosen isotropically. 
Note that, except for the cases of static (where only the direction but not the magnitude of the velocity vector can change) or absent scatterers, its choice has a negligible influence on the expansion due to the underlying Markov assumption. 

Even though we insert the various scales of our experiment as closely as possible, we emphasize that the simulation still describes a very different setting from our system. 
Nevertheless, employing the simulation reinforces the assumption that Fermi acceleration is the underlying mechanism driving our atoms to superdiffusion. 

\begin{figure}
    \centering
    \includegraphics[width=86mm]{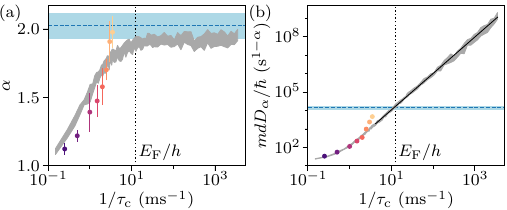}
    \caption{
    Results from extended simulation for (a)~exponent $\alpha$ and (b)~diffusion coefficient $D_\alpha$ in analogy to Fig.~\ref{fig3}, but with extended decorrelation rates up to $\decorrrate = \SI{3500}{\per\milli\second}$. 
    The graph shows the combined data from the original and extended simulations on a logarithmic \decorrrate{} axis together with the experimental results from the expansion in weak disorder, $\disorderstrength = \SI{123}{\nano K} \times \boltzmann$. 
    Blue dashed lines and areas around it show the experimental results of the disorder-free expansion. 
    The black dotted line marks $E_\mathrm{F} / h$, and the black solid line in (b) is a power-law fit to the simulation yielding an exponent of $\nu_\mathrm{FA} = 1.98 \, \pm \, 0.02$. 
    }
    \label{figS1}
\end{figure}

To investigate the maximally achievable exponent, we extend our simulation series in \decorrrate{} by three orders of magnitude, see Fig.~\ref{figS1}(a). 
As observed from our simulation data, the exponent approaches $\alpha = 2$ for large \decorrrate{}. 
As stated by Refs.~\cite{golubovic_classical_1991S, rosenbluth_comment_1992S, bouchet_minimal_2004S, aguer_classical_2010S}, $\alpha = 2$ is the expected scaling law for infinite time, which we appear to approximate closely albeit slowly. 
For any experimental system with finite size and disorder strength, the ballistic exponent $\alpha = 2$ will be the limit if enough time has passed. 
This can be intuitively understood in finite-amplitude disorder, where, after sufficient acceleration, particles will have energies above the large majority of the disorder, and the acceleration saturates. 
Effectively, there are hardly any potential features high enough to accelerate the particle further. 
On the other side, as was shown in Ref.~\cite{levi_hyper-transport_2012S}, even exponents $\alpha > 2$ beyond the ballistic regime, called hyper transport, can be achieved. 

We compare the relevant present rates and inverse timescales to estimate the efficiency with which we can increase the energy in our experimental system. 
For the maximally dynamic disorder, local disorder dynamics occur with the decorrelation rate of $\corrtime = \SI{3.5}{\per\milli\second}$~\cite{nagler_ultracold_2022S}. 
The Fermi energy corresponds to the inverse timescale $E_\mathrm{F} / h \approx \SI{12.5}{\per\milli\second}$, where $h$ is the Planck constant, \textit{i.e.}, it is larger by a factor of roughly $3.5$, see Fig.~\ref{figS1}. 
Around that rate, the exponent from the simulation begins to saturate, while the diffusion coefficient coincides with the experimentally determined $D_2$ from the disorder-free expansion. 
Alternatively, we can estimate the optimal rate $1/\tau_\mathrm{opt} \approx \SI{10.5}{\per\milli\second}$ to drive superdiffusion as efficiently as possible, as described in Ref.~\cite{volpe_brownian_2014S} for a classical two-dimensional system, yielding a similar factor. 
Thus, both comparisons indicate that we could enhance the rate of energy increase if we were to achieve higher decorrelation rates, which is technically not possible yet. 
Additionally, we find that the diffusion coefficient for decorrelation rates $1/\tau_c \geq E_\mathrm{F} / h$ follows a power law with exponent $\nu_\mathrm{FA} \approx 2$, see Fig.~\ref{figS1}(b). 
Since the kinetic energy $E_\mathrm{kin} \sim v_\mathrm{p}^2 \sim D_2$ of a particle with velocity $v_\mathrm{p}$ increases with $\decorrrate^{\nu_\mathrm{FA}} \sim v_\mathrm{s}^{\nu_\mathrm{FA}}$ (with average scatterer velocity $v_\mathrm{s}$, see Fig.~\ref{fig1}(d)), we recover $E_\mathrm{kin} \sim v_\mathrm{s}^2$, which is the expected scaling for second-order Fermi acceleration~\cite{ostrowski_diffusion_1997S, bouchet_minimal_2004S}.

\section{Localized fraction}

\begin{figure}
    \centering
    \includegraphics[width=86mm]{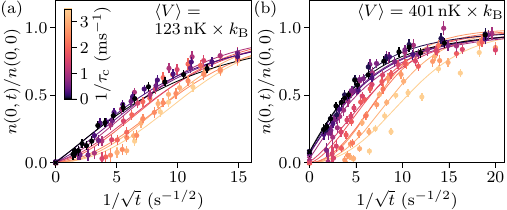}
    \caption{
    Determining the localized fraction $f_\mathrm{loc}$ from the series with (a)~weak disorder, $\disorderstrength = \SI{123}{\nano K} \times \boltzmann$, and (b)~strong disorder, $\disorderstrength = \SI{401}{\nano K} \times \boltzmann$. 
    Plotting the relative density $n(t) / n(0)$ (circles) over the square root of inverse time allows us to visualize $f_\mathrm{loc}$ as the density-axis intercept (crosses). Lines are fits with the anomalous-diffusion model Eq.~\eqref{eq:floc} where $f_\mathrm{loc}$, the value of the density for $t \rightarrow \infty$, is the only free parameter. Errors are calculated from error propagation. 
    }
    \label{figS2}
\end{figure}

The localized fraction $f_\mathrm{loc}$ estimates the infinite-time fraction of atoms that would not diffuse away due to being localized, assuming no atom losses. 
We base the determination of the localized fraction on the method reported in Ref.~\cite{jendrzejewski_three-dimensional_2012S}. 
We modified it to fit our expansion along a single dimension and implemented the full anomalous-diffusion power law as in Eq.~\eqref{eq:anomalous_diffusion}. 
More precisely, we use the model
\begin{equation}
\label{eq:floc}
    \frac{n(0, t)}{n(0, 0)} = f_\mathrm{loc} + (1 - f_\mathrm{loc}) \sqrt{\frac{\sigma^2(0)}{2 D_\alpha t^\alpha + \sigma^2(0)}}, 
\end{equation}
where we fix the diffusion exponent $\alpha$ and coefficient $D_\alpha$ to the values we extract as described in the text and use $\sigma(0) = \SI{53}{\micro\meter}$ from a Gauss fit to the trapped cloud. 
For the relative peak density $n(0, t) / n(0, 0)$, we use the above-mentioned approximation of $n(0, t) \approx w(t) - w_\mathrm{noise}$ with an additional factor of $N(0) / N(t)$ to compensate for atom losses. 
With that, $f_\mathrm{loc}$ is extracted as the only free parameter from fitting the right side of Eq.~\eqref{eq:floc}, see Fig.~\ref{figS2}. 
Further, see Ref.~\cite{barbosa_what_2023S} for more details.


\begin{thebibliography}{51}%
\makeatletter
\providecommand \@ifxundefined [1]{%
 \@ifx{#1\undefined}
}%
\providecommand \@ifnum [1]{%
 \ifnum #1\expandafter \@firstoftwo
 \else \expandafter \@secondoftwo
 \fi
}%
\providecommand \@ifx [1]{%
 \ifx #1\expandafter \@firstoftwo
 \else \expandafter \@secondoftwo
 \fi
}%
\providecommand \natexlab [1]{#1}%
\providecommand \enquote  [1]{``#1''}%
\providecommand \bibnamefont  [1]{#1}%
\providecommand \bibfnamefont [1]{#1}%
\providecommand \citenamefont [1]{#1}%
\providecommand \href@noop [0]{\@secondoftwo}%
\providecommand \href [0]{\begingroup \@sanitize@url \@href}%
\providecommand \@href[1]{\@@startlink{#1}\@@href}%
\providecommand \@@href[1]{\endgroup#1\@@endlink}%
\providecommand \@sanitize@url [0]{\catcode `\\12\catcode `\$12\catcode
  `\&12\catcode `\#12\catcode `\^12\catcode `\_12\catcode `\%12\relax}%
\providecommand \@@startlink[1]{}%
\providecommand \@@endlink[0]{}%
\providecommand \url  [0]{\begingroup\@sanitize@url \@url }%
\providecommand \@url [1]{\endgroup\@href {#1}{\urlprefix }}%
\providecommand \urlprefix  [0]{URL }%
\providecommand \Eprint [0]{\href }%
\providecommand \doibase [0]{https://doi.org/}%
\providecommand \selectlanguage [0]{\@gobble}%
\providecommand \bibinfo  [0]{\@secondoftwo}%
\providecommand \bibfield  [0]{\@secondoftwo}%
\providecommand \translation [1]{[#1]}%
\providecommand \BibitemOpen [0]{}%
\providecommand \bibitemStop [0]{}%
\providecommand \bibitemNoStop [0]{.\EOS\space}%
\providecommand \EOS [0]{\spacefactor3000\relax}%
\providecommand \BibitemShut  [1]{\csname bibitem#1\endcsname}%
\let\auto@bib@innerbib\@empty
%</preamble>
\bibitem [{\citenamefont {Einstein}(1956)}]{einstein_investigations_1956dover}%
  \BibitemOpen
  \bibfield  {author} {\bibinfo {author} {\bibfnamefont {A.}~\bibnamefont
  {Einstein}},\ }\href@noop {} {\emph {\bibinfo {title} {{Investigations} on
  the {Theory} of the {Brownian} {Movement}}}},\ edited by\ \bibinfo {editor}
  {\bibfnamefont {R.}~\bibnamefont {Fürth}}\ (\bibinfo  {publisher} {{Dover}
  {Publications}},\ \bibinfo {year} {1956})\BibitemShut {NoStop}%
\bibitem [{\citenamefont {Duplantier}(2006)}]{duplantier_brownian_2006}%
  \BibitemOpen
  \bibfield  {author} {\bibinfo {author} {\bibfnamefont {B.}~\bibnamefont
  {Duplantier}},\ }\bibfield  {title} {\bibinfo {title} {Brownian {Motion},
  “{Diverse} and {Undulating}”},\ }in\ \href
  {https://doi.org/10.1007/3-7643-7436-5_8} {\emph {\bibinfo {booktitle}
  {Einstein, 1905–2005: {Poincaré} {Seminar} 2005}}},\ \bibinfo {series and
  number} {Progress in {Mathematical} {Physics}},\ \bibinfo {editor} {edited
  by\ \bibinfo {editor} {\bibfnamefont {T.}~\bibnamefont {Damour}}, \bibinfo
  {editor} {\bibfnamefont {O.}~\bibnamefont {Darrigol}}, \bibinfo {editor}
  {\bibfnamefont {B.}~\bibnamefont {Duplantier}},\ and\ \bibinfo {editor}
  {\bibfnamefont {V.}~\bibnamefont {Rivasseau}}}\ (\bibinfo  {publisher}
  {Birkhäuser},\ \bibinfo {address} {Basel},\ \bibinfo {year} {2006})\ pp.\
  \bibinfo {pages} {201--293}\BibitemShut {NoStop}%
\bibitem [{\citenamefont {Metzler}\ and\ \citenamefont
  {Klafter}(2000)}]{metzler_random_2000}%
  \BibitemOpen
  \bibfield  {author} {\bibinfo {author} {\bibfnamefont {R.}~\bibnamefont
  {Metzler}}\ and\ \bibinfo {author} {\bibfnamefont {J.}~\bibnamefont
  {Klafter}},\ }\bibfield  {title} {\bibinfo {title} {The random walk's guide
  to anomalous diffusion: a fractional dynamics approach},\ }\href
  {https://doi.org/10.1016/S0370-1573(00)00070-3} {\bibfield  {journal}
  {\bibinfo  {journal} {Physics Reports}\ }\textbf {\bibinfo {volume} {339}},\
  \bibinfo {pages} {1} (\bibinfo {year} {2000})}\BibitemShut {NoStop}%
\bibitem [{\citenamefont {Muñoz-Gil}\ \emph {et~al.}(2021)\citenamefont
  {Muñoz-Gil}, \citenamefont {Volpe}, \citenamefont {Garcia-March},
  \citenamefont {Aghion}, \citenamefont {Argun}, \citenamefont {Hong},
  \citenamefont {Bland}, \citenamefont {Bo}, \citenamefont {Conejero},
  \citenamefont {Firbas}, \citenamefont {Garibo~i Orts}, \citenamefont
  {Gentili}, \citenamefont {Huang}, \citenamefont {Jeon}, \citenamefont
  {Kabbech}, \citenamefont {Kim}, \citenamefont {Kowalek}, \citenamefont
  {Krapf}, \citenamefont {Loch-Olszewska}, \citenamefont {Lomholt},
  \citenamefont {Masson}, \citenamefont {Meyer}, \citenamefont {Park},
  \citenamefont {Requena}, \citenamefont {Smal}, \citenamefont {Song},
  \citenamefont {Szwabiński}, \citenamefont {Thapa}, \citenamefont {Verdier},
  \citenamefont {Volpe}, \citenamefont {Widera}, \citenamefont {Lewenstein},
  \citenamefont {Metzler},\ and\ \citenamefont
  {Manzo}}]{munoz-gil_objective_2021}%
  \BibitemOpen
  \bibfield  {author} {\bibinfo {author} {\bibfnamefont {G.}~\bibnamefont
  {Muñoz-Gil}}, \bibinfo {author} {\bibfnamefont {G.}~\bibnamefont {Volpe}},
  \bibinfo {author} {\bibfnamefont {M.~A.}\ \bibnamefont {Garcia-March}},
  \bibinfo {author} {\bibfnamefont {E.}~\bibnamefont {Aghion}}, \bibinfo
  {author} {\bibfnamefont {A.}~\bibnamefont {Argun}}, \bibinfo {author}
  {\bibfnamefont {C.~B.}\ \bibnamefont {Hong}}, \bibinfo {author}
  {\bibfnamefont {T.}~\bibnamefont {Bland}}, \bibinfo {author} {\bibfnamefont
  {S.}~\bibnamefont {Bo}}, \bibinfo {author} {\bibfnamefont {J.~A.}\
  \bibnamefont {Conejero}}, \bibinfo {author} {\bibfnamefont {N.}~\bibnamefont
  {Firbas}} \textit{et al.},\ }\bibfield  {title} {\bibinfo {title} {Objective
  comparison of methods to decode anomalous diffusion},\ }\href
  {https://doi.org/10.1038/s41467-021-26320-w} {\bibfield  {journal} {\bibinfo
  {journal} {Nature Communications}\ }\textbf {\bibinfo {volume} {12}},\
  \bibinfo {pages} {6253} (\bibinfo {year} {2021})}\BibitemShut {NoStop}%
\bibitem [{\citenamefont {Uchaikin}\ and\ \citenamefont
  {Sibatov}(2017)}]{uchaikin_fractional_2017}%
  \BibitemOpen
  \bibfield  {author} {\bibinfo {author} {\bibfnamefont {V.}~\bibnamefont
  {Uchaikin}}\ and\ \bibinfo {author} {\bibfnamefont {R.}~\bibnamefont
  {Sibatov}},\ }\bibfield  {title} {\bibinfo {title} {Fractional derivatives on
  cosmic scales},\ }\href {https://doi.org/10.1016/j.chaos.2017.04.023}
  {\bibfield  {journal} {\bibinfo  {journal} {Chaos, Solitons \& Fractals}\
  }\textbf {\bibinfo {volume} {102}},\ \bibinfo {pages} {197} (\bibinfo {year}
  {2017})}\BibitemShut {NoStop}%
\bibitem [{\citenamefont {Viswanathan}\ \emph {et~al.}(1996)\citenamefont
  {Viswanathan}, \citenamefont {Afanasyev}, \citenamefont {Buldyrev},
  \citenamefont {Murphy}, \citenamefont {Prince},\ and\ \citenamefont
  {Stanley}}]{viswanathan_levy_1996}%
  \BibitemOpen
  \bibfield  {author} {\bibinfo {author} {\bibfnamefont {G.~M.}\ \bibnamefont
  {Viswanathan}}, \bibinfo {author} {\bibfnamefont {V.}~\bibnamefont
  {Afanasyev}}, \bibinfo {author} {\bibfnamefont {S.~V.}\ \bibnamefont
  {Buldyrev}}, \bibinfo {author} {\bibfnamefont {E.~J.}\ \bibnamefont
  {Murphy}}, \bibinfo {author} {\bibfnamefont {P.~A.}\ \bibnamefont {Prince}},\
  and\ \bibinfo {author} {\bibfnamefont {H.~E.}\ \bibnamefont {Stanley}},\
  }\bibfield  {title} {\bibinfo {title} {Lévy flight search patterns of
  wandering albatrosses},\ }\href {https://doi.org/10.1038/381413a0} {\bibfield
   {journal} {\bibinfo  {journal} {Nature}\ }\textbf {\bibinfo {volume}
  {381}},\ \bibinfo {pages} {413} (\bibinfo {year} {1996})}\BibitemShut
  {NoStop}%
\bibitem [{\citenamefont {Edwards}\ \emph {et~al.}(2007)\citenamefont
  {Edwards}, \citenamefont {Phillips}, \citenamefont {Watkins}, \citenamefont
  {Freeman}, \citenamefont {Murphy}, \citenamefont {Afanasyev}, \citenamefont
  {Buldyrev}, \citenamefont {da~Luz}, \citenamefont {Raposo}, \citenamefont
  {Stanley},\ and\ \citenamefont {Viswanathan}}]{edwards_revisiting_2007}%
  \BibitemOpen
  \bibfield  {author} {\bibinfo {author} {\bibfnamefont {A.~M.}\ \bibnamefont
  {Edwards}}, \bibinfo {author} {\bibfnamefont {R.~A.}\ \bibnamefont
  {Phillips}}, \bibinfo {author} {\bibfnamefont {N.~W.}\ \bibnamefont
  {Watkins}}, \bibinfo {author} {\bibfnamefont {M.~P.}\ \bibnamefont
  {Freeman}}, \bibinfo {author} {\bibfnamefont {E.~J.}\ \bibnamefont {Murphy}},
  \bibinfo {author} {\bibfnamefont {V.}~\bibnamefont {Afanasyev}}, \bibinfo
  {author} {\bibfnamefont {S.~V.}\ \bibnamefont {Buldyrev}}, \bibinfo {author}
  {\bibfnamefont {M.~G.~E.}\ \bibnamefont {da~Luz}}, \bibinfo {author}
  {\bibfnamefont {E.~P.}\ \bibnamefont {Raposo}}, \bibinfo {author}
  {\bibfnamefont {H.~E.}\ \bibnamefont {Stanley}},\ and\ \bibinfo {author}
  {\bibfnamefont {G.~M.}\ \bibnamefont {Viswanathan}},\ }\bibfield  {title}
  {\bibinfo {title} {Revisiting {Lévy} flight search patterns of wandering
  albatrosses, bumblebees and deer},\ }\href
  {https://doi.org/10.1038/nature06199} {\bibfield  {journal} {\bibinfo
  {journal} {Nature}\ }\textbf {\bibinfo {volume} {449}},\ \bibinfo {pages}
  {1044} (\bibinfo {year} {2007})}\BibitemShut {NoStop}%
\bibitem [{\citenamefont {Plerou}\ \emph {et~al.}(2000)\citenamefont {Plerou},
  \citenamefont {Gopikrishnan}, \citenamefont {Nunes~Amaral}, \citenamefont
  {Gabaix},\ and\ \citenamefont {Eugene~Stanley}}]{plerou_economic_2000}%
  \BibitemOpen
  \bibfield  {author} {\bibinfo {author} {\bibfnamefont {V.}~\bibnamefont
  {Plerou}}, \bibinfo {author} {\bibfnamefont {P.}~\bibnamefont
  {Gopikrishnan}}, \bibinfo {author} {\bibfnamefont {L.~A.}\ \bibnamefont
  {Nunes~Amaral}}, \bibinfo {author} {\bibfnamefont {X.}~\bibnamefont
  {Gabaix}},\ and\ \bibinfo {author} {\bibfnamefont {H.}~\bibnamefont
  {Eugene~Stanley}},\ }\bibfield  {title} {\bibinfo {title} {Economic
  fluctuations and anomalous diffusion},\ }\href
  {https://doi.org/10.1103/PhysRevE.62.R3023} {\bibfield  {journal} {\bibinfo
  {journal} {Physical Review E}\ }\textbf {\bibinfo {volume} {62}},\ \bibinfo
  {pages} {R3023} (\bibinfo {year} {2000})}\BibitemShut {NoStop}%
\bibitem [{\citenamefont {Balescu}(1995)}]{balescu_anomalous_1995}%
  \BibitemOpen
  \bibfield  {author} {\bibinfo {author} {\bibfnamefont {R.}~\bibnamefont
  {Balescu}},\ }\bibfield  {title} {\bibinfo {title} {Anomalous transport in
  turbulent plasmas and continuous time random walks},\ }\href
  {https://doi.org/10.1103/PhysRevE.51.4807} {\bibfield  {journal} {\bibinfo
  {journal} {Physical Review E}\ }\textbf {\bibinfo {volume} {51}},\ \bibinfo
  {pages} {4807} (\bibinfo {year} {1995})}\BibitemShut {NoStop}%
\bibitem [{\citenamefont {Di~Pierro}\ \emph {et~al.}(2018)\citenamefont
  {Di~Pierro}, \citenamefont {Potoyan}, \citenamefont {Wolynes},\ and\
  \citenamefont {Onuchic}}]{di_pierro_anomalous_2018}%
  \BibitemOpen
  \bibfield  {author} {\bibinfo {author} {\bibfnamefont {M.}~\bibnamefont
  {Di~Pierro}}, \bibinfo {author} {\bibfnamefont {D.~A.}\ \bibnamefont
  {Potoyan}}, \bibinfo {author} {\bibfnamefont {P.~G.}\ \bibnamefont
  {Wolynes}},\ and\ \bibinfo {author} {\bibfnamefont {J.~N.}\ \bibnamefont
  {Onuchic}},\ }\bibfield  {title} {\bibinfo {title} {Anomalous diffusion,
  spatial coherence, and viscoelasticity from the energy landscape of human
  chromosomes},\ }\href {https://doi.org/10.1073/pnas.1806297115} {\bibfield
  {journal} {\bibinfo  {journal} {Proceedings of the National Academy of
  Sciences}\ }\textbf {\bibinfo {volume} {115}},\ \bibinfo {pages} {7753}
  (\bibinfo {year} {2018})}\BibitemShut {NoStop}%
\bibitem [{\citenamefont {Anderson}(1958)}]{anderson_absence_1958}%
  \BibitemOpen
  \bibfield  {author} {\bibinfo {author} {\bibfnamefont {P.~W.}\ \bibnamefont
  {Anderson}},\ }\bibfield  {title} {\bibinfo {title} {Absence of {Diffusion}
  in {Certain} {Random} {Lattices}},\ }\href
  {https://doi.org/10.1103/PhysRev.109.1492} {\bibfield  {journal} {\bibinfo
  {journal} {Physical Review}\ }\textbf {\bibinfo {volume} {109}},\ \bibinfo
  {pages} {1492} (\bibinfo {year} {1958})}\BibitemShut {NoStop}%
\bibitem [{\citenamefont {Abrahams}(2010)}]{AndersonLocalization}%
  \BibitemOpen
  \bibinfo {editor} {\bibfnamefont {E.}~\bibnamefont {Abrahams}},\ ed.,\ \href
  {https://doi.org/10.1142/7663} {\emph {\bibinfo {title} {50 Years of Anderson
  Localization}}}\ (\bibinfo  {publisher} {World Scientific},\ \bibinfo
  {address} {Singapore},\ \bibinfo {year} {2010})\BibitemShut {NoStop}%
\bibitem [{\citenamefont {Weaver}(1990)}]{Weaver1990}%
  \BibitemOpen
  \bibfield  {author} {\bibinfo {author} {\bibfnamefont {R.}~\bibnamefont
  {Weaver}},\ }\bibfield  {title} {\bibinfo {title} {Anderson localization of
  ultrasound},\ }\href
  {https://doi.org/https://doi.org/10.1016/0165-2125(90)90034-2} {\bibfield
  {journal} {\bibinfo  {journal} {Wave Motion}\ }\textbf {\bibinfo {volume}
  {12}},\ \bibinfo {pages} {129} (\bibinfo {year} {1990})}\BibitemShut
  {NoStop}%
\bibitem [{\citenamefont {Hu}\ \emph {et~al.}(2008)\citenamefont {Hu},
  \citenamefont {Strybulevych}, \citenamefont {Page}, \citenamefont
  {Skipetrov},\ and\ \citenamefont {van Tiggelen}}]{Hu2008}%
  \BibitemOpen
  \bibfield  {author} {\bibinfo {author} {\bibfnamefont {H.}~\bibnamefont
  {Hu}}, \bibinfo {author} {\bibfnamefont {A.}~\bibnamefont {Strybulevych}},
  \bibinfo {author} {\bibfnamefont {J.~H.}\ \bibnamefont {Page}}, \bibinfo
  {author} {\bibfnamefont {S.~E.}\ \bibnamefont {Skipetrov}},\ and\ \bibinfo
  {author} {\bibfnamefont {B.~A.}\ \bibnamefont {van Tiggelen}},\ }\bibfield
  {title} {\bibinfo {title} {Localization of ultrasound in a three-dimensional
  elastic network},\ }\href {https://doi.org/10.1038/nphys1101} {\bibfield
  {journal} {\bibinfo  {journal} {Nature Physics}\ }\textbf {\bibinfo {volume}
  {4}},\ \bibinfo {pages} {945} (\bibinfo {year} {2008})}\BibitemShut {NoStop}%
\bibitem [{\citenamefont {Dalichaouch}\ \emph {et~al.}(1991)\citenamefont
  {Dalichaouch}, \citenamefont {Armstrong}, \citenamefont {Schultz},
  \citenamefont {Platzman},\ and\ \citenamefont {McCall}}]{Dalichaouch1991}%
  \BibitemOpen
  \bibfield  {author} {\bibinfo {author} {\bibfnamefont {R.}~\bibnamefont
  {Dalichaouch}}, \bibinfo {author} {\bibfnamefont {J.~P.}\ \bibnamefont
  {Armstrong}}, \bibinfo {author} {\bibfnamefont {S.}~\bibnamefont {Schultz}},
  \bibinfo {author} {\bibfnamefont {P.~M.}\ \bibnamefont {Platzman}},\ and\
  \bibinfo {author} {\bibfnamefont {S.~L.}\ \bibnamefont {McCall}},\ }\bibfield
   {title} {\bibinfo {title} {Microwave localization by two-dimensional random
  scattering},\ }\href {https://doi.org/10.1038/354053a0} {\bibfield  {journal}
  {\bibinfo  {journal} {Nature}\ }\textbf {\bibinfo {volume} {354}},\ \bibinfo
  {pages} {53} (\bibinfo {year} {1991})}\BibitemShut {NoStop}%
\bibitem [{\citenamefont {Wiersma}\ \emph {et~al.}(1997)\citenamefont
  {Wiersma}, \citenamefont {Bartolini}, \citenamefont {Lagendijk},\ and\
  \citenamefont {Righini}}]{Wiersma1997}%
  \BibitemOpen
  \bibfield  {author} {\bibinfo {author} {\bibfnamefont {D.~S.}\ \bibnamefont
  {Wiersma}}, \bibinfo {author} {\bibfnamefont {P.}~\bibnamefont {Bartolini}},
  \bibinfo {author} {\bibfnamefont {A.}~\bibnamefont {Lagendijk}},\ and\
  \bibinfo {author} {\bibfnamefont {R.}~\bibnamefont {Righini}},\ }\bibfield
  {title} {\bibinfo {title} {Localization of light in a disordered medium},\
  }\href {https://doi.org/10.1038/37757} {\bibfield  {journal} {\bibinfo
  {journal} {Nature}\ }\textbf {\bibinfo {volume} {390}},\ \bibinfo {pages}
  {671} (\bibinfo {year} {1997})}\BibitemShut {NoStop}%
\bibitem [{\citenamefont {Scheffold}\ \emph {et~al.}(1999)\citenamefont
  {Scheffold}, \citenamefont {Lenke}, \citenamefont {Tweer},\ and\
  \citenamefont {Maret}}]{Scheffold1999}%
  \BibitemOpen
  \bibfield  {author} {\bibinfo {author} {\bibfnamefont {F.}~\bibnamefont
  {Scheffold}}, \bibinfo {author} {\bibfnamefont {R.}~\bibnamefont {Lenke}},
  \bibinfo {author} {\bibfnamefont {R.}~\bibnamefont {Tweer}},\ and\ \bibinfo
  {author} {\bibfnamefont {G.}~\bibnamefont {Maret}},\ }\bibfield  {title}
  {\bibinfo {title} {Localization or classical diffusion of light?},\ }\href
  {https://doi.org/10.1038/18347} {\bibfield  {journal} {\bibinfo  {journal}
  {Nature}\ }\textbf {\bibinfo {volume} {398}},\ \bibinfo {pages} {206}
  (\bibinfo {year} {1999})}\BibitemShut {NoStop}%
\bibitem [{\citenamefont {Schwartz}\ \emph {et~al.}(2007)\citenamefont
  {Schwartz}, \citenamefont {Bartal}, \citenamefont {Fishman},\ and\
  \citenamefont {Segev}}]{schwartz_transport_2007}%
  \BibitemOpen
  \bibfield  {author} {\bibinfo {author} {\bibfnamefont {T.}~\bibnamefont
  {Schwartz}}, \bibinfo {author} {\bibfnamefont {G.}~\bibnamefont {Bartal}},
  \bibinfo {author} {\bibfnamefont {S.}~\bibnamefont {Fishman}},\ and\ \bibinfo
  {author} {\bibfnamefont {M.}~\bibnamefont {Segev}},\ }\bibfield  {title}
  {\bibinfo {title} {Transport and {Anderson} localization in disordered
  two-dimensional photonic lattices},\ }\href
  {https://doi.org/10.1038/nature05623} {\bibfield  {journal} {\bibinfo
  {journal} {Nature}\ }\textbf {\bibinfo {volume} {446}},\ \bibinfo {pages}
  {52} (\bibinfo {year} {2007})}\BibitemShut {NoStop}%
\bibitem [{\citenamefont {Mafi}\ and\ \citenamefont
  {Ballato}(2021)}]{Mafi2021}%
  \BibitemOpen
  \bibfield  {author} {\bibinfo {author} {\bibfnamefont {A.}~\bibnamefont
  {Mafi}}\ and\ \bibinfo {author} {\bibfnamefont {J.}~\bibnamefont {Ballato}},\
  }\bibfield  {title} {\bibinfo {title} {{Review} of a {Decade} of {Research}
  on {Disordered} {Anderson} {Localizing} {Optical} {Fibers}},\ }\href
  {https://www.frontiersin.org/articles/10.3389/fphy.2021.736774} {\bibfield
  {journal} {\bibinfo  {journal} {Frontiers in Physics}\ }\textbf {\bibinfo
  {volume} {9}} (\bibinfo {year} {2021})}\BibitemShut {NoStop}%
\bibitem [{\citenamefont {Billy}\ \emph {et~al.}(2008)\citenamefont {Billy},
  \citenamefont {Josse}, \citenamefont {Zuo}, \citenamefont {Bernard},
  \citenamefont {Hambrecht}, \citenamefont {Lugan}, \citenamefont {Clément},
  \citenamefont {Sanchez-Palencia}, \citenamefont {Bouyer},\ and\ \citenamefont
  {Aspect}}]{billy_direct_2008}%
  \BibitemOpen
  \bibfield  {author} {\bibinfo {author} {\bibfnamefont {J.}~\bibnamefont
  {Billy}}, \bibinfo {author} {\bibfnamefont {V.}~\bibnamefont {Josse}},
  \bibinfo {author} {\bibfnamefont {Z.}~\bibnamefont {Zuo}}, \bibinfo {author}
  {\bibfnamefont {A.}~\bibnamefont {Bernard}}, \bibinfo {author} {\bibfnamefont
  {B.}~\bibnamefont {Hambrecht}}, \bibinfo {author} {\bibfnamefont
  {P.}~\bibnamefont {Lugan}}, \bibinfo {author} {\bibfnamefont
  {D.}~\bibnamefont {Clément}}, \bibinfo {author} {\bibfnamefont
  {L.}~\bibnamefont {Sanchez-Palencia}}, \bibinfo {author} {\bibfnamefont
  {P.}~\bibnamefont {Bouyer}},\ and\ \bibinfo {author} {\bibfnamefont
  {A.}~\bibnamefont {Aspect}},\ }\bibfield  {title} {\bibinfo {title} {Direct
  observation of {Anderson} localization of matter waves in a controlled
  disorder},\ }\href {https://doi.org/10.1038/nature07000} {\bibfield
  {journal} {\bibinfo  {journal} {Nature}\ }\textbf {\bibinfo {volume} {453}},\
  \bibinfo {pages} {891} (\bibinfo {year} {2008})}\BibitemShut {NoStop}%
\bibitem [{\citenamefont {Roati}\ \emph {et~al.}(2008)\citenamefont {Roati},
  \citenamefont {D’Errico}, \citenamefont {Fallani}, \citenamefont {Fattori},
  \citenamefont {Fort}, \citenamefont {Zaccanti}, \citenamefont {Modugno},
  \citenamefont {Modugno},\ and\ \citenamefont
  {Inguscio}}]{roati_anderson_2008}%
  \BibitemOpen
  \bibfield  {author} {\bibinfo {author} {\bibfnamefont {G.}~\bibnamefont
  {Roati}}, \bibinfo {author} {\bibfnamefont {C.}~\bibnamefont {D’Errico}},
  \bibinfo {author} {\bibfnamefont {L.}~\bibnamefont {Fallani}}, \bibinfo
  {author} {\bibfnamefont {M.}~\bibnamefont {Fattori}}, \bibinfo {author}
  {\bibfnamefont {C.}~\bibnamefont {Fort}}, \bibinfo {author} {\bibfnamefont
  {M.}~\bibnamefont {Zaccanti}}, \bibinfo {author} {\bibfnamefont
  {G.}~\bibnamefont {Modugno}}, \bibinfo {author} {\bibfnamefont
  {M.}~\bibnamefont {Modugno}},\ and\ \bibinfo {author} {\bibfnamefont
  {M.}~\bibnamefont {Inguscio}},\ }\bibfield  {title} {\bibinfo {title}
  {Anderson localization of a non-interacting {Bose}–{Einstein} condensate},\
  }\href {https://doi.org/10.1038/nature07071} {\bibfield  {journal} {\bibinfo
  {journal} {Nature}\ }\textbf {\bibinfo {volume} {453}},\ \bibinfo {pages}
  {895} (\bibinfo {year} {2008})}\BibitemShut {NoStop}%
\bibitem [{\citenamefont {Jendrzejewski}\ \emph {et~al.}(2012)\citenamefont
  {Jendrzejewski}, \citenamefont {Bernard}, \citenamefont {Müller},
  \citenamefont {Cheinet}, \citenamefont {Josse}, \citenamefont {Piraud},
  \citenamefont {Pezzé}, \citenamefont {Sanchez-Palencia}, \citenamefont
  {Aspect},\ and\ \citenamefont
  {Bouyer}}]{jendrzejewski_three-dimensional_2012}%
  \BibitemOpen
  \bibfield  {author} {\bibinfo {author} {\bibfnamefont {F.}~\bibnamefont
  {Jendrzejewski}}, \bibinfo {author} {\bibfnamefont {A.}~\bibnamefont
  {Bernard}}, \bibinfo {author} {\bibfnamefont {K.}~\bibnamefont {Müller}},
  \bibinfo {author} {\bibfnamefont {P.}~\bibnamefont {Cheinet}}, \bibinfo
  {author} {\bibfnamefont {V.}~\bibnamefont {Josse}}, \bibinfo {author}
  {\bibfnamefont {M.}~\bibnamefont {Piraud}}, \bibinfo {author} {\bibfnamefont
  {L.}~\bibnamefont {Pezzé}}, \bibinfo {author} {\bibfnamefont
  {L.}~\bibnamefont {Sanchez-Palencia}}, \bibinfo {author} {\bibfnamefont
  {A.}~\bibnamefont {Aspect}},\ and\ \bibinfo {author} {\bibfnamefont
  {P.}~\bibnamefont {Bouyer}},\ }\bibfield  {title} {\bibinfo {title}
  {Three-dimensional localization of ultracold atoms in an optical disordered
  potential},\ }\href {https://doi.org/10.1038/nphys2256} {\bibfield  {journal}
  {\bibinfo  {journal} {Nature Physics}\ }\textbf {\bibinfo {volume} {8}},\
  \bibinfo {pages} {398} (\bibinfo {year} {2012})}\BibitemShut {NoStop}%
\bibitem [{\citenamefont {Kondov}\ \emph {et~al.}(2011)\citenamefont {Kondov},
  \citenamefont {McGehee}, \citenamefont {Zirbel},\ and\ \citenamefont
  {DeMarco}}]{kondov_three-dimensional_2011}%
  \BibitemOpen
  \bibfield  {author} {\bibinfo {author} {\bibfnamefont {S.~S.}\ \bibnamefont
  {Kondov}}, \bibinfo {author} {\bibfnamefont {W.~R.}\ \bibnamefont {McGehee}},
  \bibinfo {author} {\bibfnamefont {J.~J.}\ \bibnamefont {Zirbel}},\ and\
  \bibinfo {author} {\bibfnamefont {B.}~\bibnamefont {DeMarco}},\ }\bibfield
  {title} {\bibinfo {title} {Three-{Dimensional} {Anderson} {Localization} of
  {Ultracold} {Matter}},\ }\href {https://doi.org/10.1126/science.1209019}
  {\bibfield  {journal} {\bibinfo  {journal} {Science}\ }\textbf {\bibinfo
  {volume} {334}},\ \bibinfo {pages} {66} (\bibinfo {year} {2011})}\BibitemShut
  {NoStop}%
\bibitem [{\citenamefont {Shapiro}(2012)}]{shapiro_cold_2012}%
  \BibitemOpen
  \bibfield  {author} {\bibinfo {author} {\bibfnamefont {B.}~\bibnamefont
  {Shapiro}},\ }\bibfield  {title} {\bibinfo {title} {Cold atoms in the
  presence of disorder},\ }\href
  {https://doi.org/10.1088/1751-8113/45/14/143001} {\bibfield  {journal}
  {\bibinfo  {journal} {Journal of Physics A: Mathematical and Theoretical}\
  }\textbf {\bibinfo {volume} {45}},\ \bibinfo {pages} {143001} (\bibinfo
  {year} {2012})}\BibitemShut {NoStop}%
\bibitem [{\citenamefont {Evensky}\ \emph {et~al.}(1990)\citenamefont
  {Evensky}, \citenamefont {Scalettar},\ and\ \citenamefont
  {Wolynes}}]{evensky_localization_1990}%
  \BibitemOpen
  \bibfield  {author} {\bibinfo {author} {\bibfnamefont {D.~A.}\ \bibnamefont
  {Evensky}}, \bibinfo {author} {\bibfnamefont {R.~T.}\ \bibnamefont
  {Scalettar}},\ and\ \bibinfo {author} {\bibfnamefont {P.~G.}\ \bibnamefont
  {Wolynes}},\ }\bibfield  {title} {\bibinfo {title} {Localization and
  dephasing effects in a time-dependent {Anderson} {Hamiltonian}},\ }\href
  {https://doi.org/10.1021/j100366a027} {\bibfield  {journal} {\bibinfo
  {journal} {The Journal of Physical Chemistry}\ }\textbf {\bibinfo {volume}
  {94}},\ \bibinfo {pages} {1149} (\bibinfo {year} {1990})}\BibitemShut
  {NoStop}%
\bibitem [{\citenamefont {Lorenzo}\ \emph {et~al.}(2018)\citenamefont
  {Lorenzo}, \citenamefont {Apollaro}, \citenamefont {Palma}, \citenamefont
  {Nandkishore}, \citenamefont {Silva},\ and\ \citenamefont
  {Marino}}]{lorenzo_remnants_2018}%
  \BibitemOpen
  \bibfield  {author} {\bibinfo {author} {\bibfnamefont {S.}~\bibnamefont
  {Lorenzo}}, \bibinfo {author} {\bibfnamefont {T.}~\bibnamefont {Apollaro}},
  \bibinfo {author} {\bibfnamefont {G.~M.}\ \bibnamefont {Palma}}, \bibinfo
  {author} {\bibfnamefont {R.}~\bibnamefont {Nandkishore}}, \bibinfo {author}
  {\bibfnamefont {A.}~\bibnamefont {Silva}},\ and\ \bibinfo {author}
  {\bibfnamefont {J.}~\bibnamefont {Marino}},\ }\bibfield  {title} {\bibinfo
  {title} {Remnants of {Anderson} localization in prethermalization induced by
  white noise},\ }\href {https://doi.org/10.1103/PhysRevB.98.054302} {\bibfield
   {journal} {\bibinfo  {journal} {Physical Review B}\ }\textbf {\bibinfo
  {volume} {98}},\ \bibinfo {pages} {054302} (\bibinfo {year}
  {2018})}\BibitemShut {NoStop}%
\bibitem [{\citenamefont {Levi}\ \emph {et~al.}(2012)\citenamefont {Levi},
  \citenamefont {Krivolapov}, \citenamefont {Fishman},\ and\ \citenamefont
  {Segev}}]{levi_hyper-transport_2012}%
  \BibitemOpen
  \bibfield  {author} {\bibinfo {author} {\bibfnamefont {L.}~\bibnamefont
  {Levi}}, \bibinfo {author} {\bibfnamefont {Y.}~\bibnamefont {Krivolapov}},
  \bibinfo {author} {\bibfnamefont {S.}~\bibnamefont {Fishman}},\ and\ \bibinfo
  {author} {\bibfnamefont {M.}~\bibnamefont {Segev}},\ }\bibfield  {title}
  {\bibinfo {title} {Hyper-transport of light and stochastic acceleration by
  evolving disorder},\ }\href {https://doi.org/10.1038/nphys2463} {\bibfield
  {journal} {\bibinfo  {journal} {Nature Physics}\ }\textbf {\bibinfo {volume}
  {8}},\ \bibinfo {pages} {912} (\bibinfo {year} {2012})}\BibitemShut {NoStop}%
\bibitem [{\citenamefont {Fermi}(1949)}]{fermi_origin_1949}%
  \BibitemOpen
  \bibfield  {author} {\bibinfo {author} {\bibfnamefont {E.}~\bibnamefont
  {Fermi}},\ }\bibfield  {title} {\bibinfo {title} {On the {Origin} of the
  {Cosmic} {Radiation}},\ }\href {https://doi.org/10.1103/PhysRev.75.1169}
  {\bibfield  {journal} {\bibinfo  {journal} {Physical Review}\ }\textbf
  {\bibinfo {volume} {75}},\ \bibinfo {pages} {1169} (\bibinfo {year}
  {1949})}\BibitemShut {NoStop}%
\bibitem [{\citenamefont {Sturrock}(1966)}]{sturrock_model_1966}%
  \BibitemOpen
  \bibfield  {author} {\bibinfo {author} {\bibfnamefont {P.~A.}\ \bibnamefont
  {Sturrock}},\ }\bibfield  {title} {\bibinfo {title} {Model of the
  {High}-{Energy} {Phase} of {Solar} {Flares}},\ }\href
  {https://doi.org/10.1038/211695a0} {\bibfield  {journal} {\bibinfo  {journal}
  {Nature}\ }\textbf {\bibinfo {volume} {211}},\ \bibinfo {pages} {695}
  (\bibinfo {year} {1966})}\BibitemShut {NoStop}%
\bibitem [{\citenamefont {Ostrowski}\ and\ \citenamefont
  {Siemieniec-Ozi\c{e}bło}(1997)}]{ostrowski_diffusion_1997}%
  \BibitemOpen
  \bibfield  {author} {\bibinfo {author} {\bibfnamefont {M.}~\bibnamefont
  {Ostrowski}}\ and\ \bibinfo {author} {\bibfnamefont {G.}~\bibnamefont
  {Siemieniec-Ozi\c{e}bło}},\ }\bibfield  {title} {\bibinfo {title} {Diffusion
  in momentum space as a picture of second-order {Fermi} acceleration},\ }\href
  {https://doi.org/10.1016/S0927-6505(96)00061-8} {\bibfield  {journal}
  {\bibinfo  {journal} {Astroparticle Physics}\ }\textbf {\bibinfo {volume}
  {6}},\ \bibinfo {pages} {271} (\bibinfo {year} {1997})}\BibitemShut {NoStop}%
\bibitem [{\citenamefont {Mertsch}\ and\ \citenamefont
  {Sarkar}(2011)}]{mertsch_fermi_2011}%
  \BibitemOpen
  \bibfield  {author} {\bibinfo {author} {\bibfnamefont {P.}~\bibnamefont
  {Mertsch}}\ and\ \bibinfo {author} {\bibfnamefont {S.}~\bibnamefont
  {Sarkar}},\ }\bibfield  {title} {\bibinfo {title} {Fermi {Gamma}-{Ray}
  “{Bubbles}” from {Stochastic} {Acceleration} of {Electrons}},\ }\href
  {https://doi.org/10.1103/PhysRevLett.107.091101} {\bibfield  {journal}
  {\bibinfo  {journal} {Physical Review Letters}\ }\textbf {\bibinfo {volume}
  {107}},\ \bibinfo {pages} {091101} (\bibinfo {year} {2011})}\BibitemShut
  {NoStop}%
\bibitem [{\citenamefont {Golubović}\ \emph {et~al.}(1991)\citenamefont
  {Golubović}, \citenamefont {Feng},\ and\ \citenamefont
  {Zeng}}]{golubovic_classical_1991}%
  \BibitemOpen
  \bibfield  {author} {\bibinfo {author} {\bibfnamefont {L.}~\bibnamefont
  {Golubović}}, \bibinfo {author} {\bibfnamefont {S.}~\bibnamefont {Feng}},\
  and\ \bibinfo {author} {\bibfnamefont {F.-A.}\ \bibnamefont {Zeng}},\
  }\bibfield  {title} {\bibinfo {title} {Classical and quantum superdiffusion
  in a time-dependent random potential},\ }\href
  {https://doi.org/10.1103/PhysRevLett.67.2115} {\bibfield  {journal} {\bibinfo
   {journal} {Physical Review Letters}\ }\textbf {\bibinfo {volume} {67}},\
  \bibinfo {pages} {2115} (\bibinfo {year} {1991})}\BibitemShut {NoStop}%
\bibitem [{\citenamefont {Rosenbluth}(1992)}]{rosenbluth_comment_1992}%
  \BibitemOpen
  \bibfield  {author} {\bibinfo {author} {\bibfnamefont {M.~N.}\ \bibnamefont
  {Rosenbluth}},\ }\bibfield  {title} {\bibinfo {title} {Comment on
  ``{Classical} and quantum superdiffusion in a time-dependent random
  potential''},\ }\href {https://doi.org/10.1103/PhysRevLett.69.1831}
  {\bibfield  {journal} {\bibinfo  {journal} {Physical Review Letters}\
  }\textbf {\bibinfo {volume} {69}},\ \bibinfo {pages} {1831} (\bibinfo {year}
  {1992})}\BibitemShut {NoStop}%
\bibitem [{\citenamefont {Aguer}\ \emph {et~al.}(2010)\citenamefont {Aguer},
  \citenamefont {De~Bièvre}, \citenamefont {Lafitte},\ and\ \citenamefont
  {Parris}}]{aguer_classical_2010}%
  \BibitemOpen
  \bibfield  {author} {\bibinfo {author} {\bibfnamefont {B.}~\bibnamefont
  {Aguer}}, \bibinfo {author} {\bibfnamefont {S.}~\bibnamefont {De~Bièvre}},
  \bibinfo {author} {\bibfnamefont {P.}~\bibnamefont {Lafitte}},\ and\ \bibinfo
  {author} {\bibfnamefont {P.~E.}\ \bibnamefont {Parris}},\ }\bibfield  {title}
  {\bibinfo {title} {Classical {Motion} in {Force} {Fields} with {Short}
  {Range} {Correlations}},\ }\href {https://doi.org/10.1007/s10955-009-9898-7}
  {\bibfield  {journal} {\bibinfo  {journal} {Journal of Statistical Physics}\
  }\textbf {\bibinfo {volume} {138}},\ \bibinfo {pages} {780} (\bibinfo {year}
  {2010})}\BibitemShut {NoStop}%
\bibitem [{\citenamefont {Volpe}\ \emph {et~al.}(2014)\citenamefont {Volpe},
  \citenamefont {Volpe},\ and\ \citenamefont {Gigan}}]{volpe_brownian_2014}%
  \BibitemOpen
  \bibfield  {author} {\bibinfo {author} {\bibfnamefont {G.}~\bibnamefont
  {Volpe}}, \bibinfo {author} {\bibfnamefont {G.}~\bibnamefont {Volpe}},\ and\
  \bibinfo {author} {\bibfnamefont {S.}~\bibnamefont {Gigan}},\ }\bibfield
  {title} {\bibinfo {title} {Brownian {Motion} in a {Speckle} {Light} {Field}:
  {Tunable} {Anomalous} {Diffusion} and {Selective} {Optical} {Manipulation}},\
  }\href {https://doi.org/10.1038/srep03936} {\bibfield  {journal} {\bibinfo
  {journal} {Scientific Reports}\ }\textbf {\bibinfo {volume} {4}},\ \bibinfo
  {pages} {3936} (\bibinfo {year} {2014})}\BibitemShut {NoStop}%
\bibitem [{\citenamefont {Ulam}(1961)}]{ulam_statistical_1961}%
  \BibitemOpen
  \bibfield  {author} {\bibinfo {author} {\bibfnamefont {S.~M.}\ \bibnamefont
  {Ulam}},\ }\bibfield  {title} {\bibinfo {title} {{On} {some} {statistical}
  {properties} {of} {dynamical} {systems}},\ }in\ \href
  {https://doi.org/10.1525/9780520323438-017} {\emph {\bibinfo {booktitle}
  {Contributions to {Astronomy}, {Meteorology}, and {Physics}}}},\ \bibinfo
  {editor} {edited by\ \bibinfo {editor} {\bibfnamefont {J.}~\bibnamefont
  {Neyman}}}\ (\bibinfo  {publisher} {University of California Press},\
  \bibinfo {year} {1961})\ pp.\ \bibinfo {pages} {315--320}\BibitemShut
  {NoStop}%
\bibitem [{\citenamefont {Lichtenberg}\ and\ \citenamefont
  {Lieberman}(1983)}]{lichtenberg_regular_1983}%
  \BibitemOpen
  \bibfield  {author} {\bibinfo {author} {\bibfnamefont {A.~J.}\ \bibnamefont
  {Lichtenberg}}\ and\ \bibinfo {author} {\bibfnamefont {M.~A.}\ \bibnamefont
  {Lieberman}},\ }\href {https://doi.org/10.1007/978-1-4757-4257-2} {\emph
  {\bibinfo {title} {Regular and {Stochastic} {Motion}}}},\ edited by\ \bibinfo
  {editor} {\bibfnamefont {F.}~\bibnamefont {John}}, \bibinfo {editor}
  {\bibfnamefont {J.~E.}\ \bibnamefont {Marsden}},\ and\ \bibinfo {editor}
  {\bibfnamefont {L.}~\bibnamefont {Sirovich}},\ \bibinfo {series} {Applied
  {Mathematical} {Sciences}}, Vol.~\bibinfo {volume} {38}\ (\bibinfo
  {publisher} {Springer},\ \bibinfo {address} {New York, NY},\ \bibinfo {year}
  {1983})\BibitemShut {NoStop}%
\bibitem [{\citenamefont {José}\ and\ \citenamefont
  {Cordery}(1986)}]{jose_study_1986}%
  \BibitemOpen
  \bibfield  {author} {\bibinfo {author} {\bibfnamefont {J.~V.}\ \bibnamefont
  {José}}\ and\ \bibinfo {author} {\bibfnamefont {R.}~\bibnamefont
  {Cordery}},\ }\bibfield  {title} {\bibinfo {title} {Study of a quantum
  fermi-acceleration model},\ }\href
  {https://doi.org/10.1103/PhysRevLett.56.290} {\bibfield  {journal} {\bibinfo
  {journal} {Physical Review Letters}\ }\textbf {\bibinfo {volume} {56}},\
  \bibinfo {pages} {290} (\bibinfo {year} {1986})}\BibitemShut {NoStop}%
\bibitem [{\citenamefont {Seba}(1990)}]{seba_quantum_1990}%
  \BibitemOpen
  \bibfield  {author} {\bibinfo {author} {\bibfnamefont {P.}~\bibnamefont
  {Seba}},\ }\bibfield  {title} {\bibinfo {title} {Quantum chaos in the
  {Fermi}-accelerator model},\ }\href
  {https://doi.org/10.1103/PhysRevA.41.2306} {\bibfield  {journal} {\bibinfo
  {journal} {Physical Review A}\ }\textbf {\bibinfo {volume} {41}},\ \bibinfo
  {pages} {2306} (\bibinfo {year} {1990})}\BibitemShut {NoStop}%
\bibitem [{\citenamefont {Bouchet}\ \emph {et~al.}(2004)\citenamefont
  {Bouchet}, \citenamefont {Cecconi},\ and\ \citenamefont
  {Vulpiani}}]{bouchet_minimal_2004}%
  \BibitemOpen
  \bibfield  {author} {\bibinfo {author} {\bibfnamefont {F.}~\bibnamefont
  {Bouchet}}, \bibinfo {author} {\bibfnamefont {F.}~\bibnamefont {Cecconi}},\
  and\ \bibinfo {author} {\bibfnamefont {A.}~\bibnamefont {Vulpiani}},\
  }\bibfield  {title} {\bibinfo {title} {Minimal {Stochastic} {Model} for
  {Fermi}'s {Acceleration}},\ }\href
  {https://doi.org/10.1103/PhysRevLett.92.040601} {\bibfield  {journal}
  {\bibinfo  {journal} {Physical Review Letters}\ }\textbf {\bibinfo {volume}
  {92}},\ \bibinfo {pages} {040601} (\bibinfo {year} {2004})}\BibitemShut
  {NoStop}%
\bibitem [{\citenamefont {Nagler}\ \emph
  {et~al.}(2022{\natexlab{a}})\citenamefont {Nagler}, \citenamefont {Will},
  \citenamefont {Hiebel}, \citenamefont {Barbosa}, \citenamefont {Koch},
  \citenamefont {Fleischhauer},\ and\ \citenamefont
  {Widera}}]{nagler_ultracold_2022}%
  \BibitemOpen
  \bibfield  {author} {\bibinfo {author} {\bibfnamefont {B.}~\bibnamefont
  {Nagler}}, \bibinfo {author} {\bibfnamefont {M.}~\bibnamefont {Will}},
  \bibinfo {author} {\bibfnamefont {S.}~\bibnamefont {Hiebel}}, \bibinfo
  {author} {\bibfnamefont {S.}~\bibnamefont {Barbosa}}, \bibinfo {author}
  {\bibfnamefont {J.}~\bibnamefont {Koch}}, \bibinfo {author} {\bibfnamefont
  {M.}~\bibnamefont {Fleischhauer}},\ and\ \bibinfo {author} {\bibfnamefont
  {A.}~\bibnamefont {Widera}},\ }\bibfield  {title} {\bibinfo {title}
  {Ultracold {Bose} {Gases} in {Dynamic} {Disorder} with {Tunable}
  {Correlation} {Time}},\ }\href
  {https://doi.org/10.1103/PhysRevLett.128.233601} {\bibfield  {journal}
  {\bibinfo  {journal} {Physical Review Letters}\ }\textbf {\bibinfo {volume}
  {128}},\ \bibinfo {pages} {233601} (\bibinfo {year}
  {2022}{\natexlab{a}})}\BibitemShut {NoStop}%
\bibitem [{\citenamefont {Hiebel}\ \emph {et~al.}(2024)\citenamefont {Hiebel},
  \citenamefont {Nagler}, \citenamefont {Barbosa}, \citenamefont {Koch},\ and\
  \citenamefont {Widera}}]{hiebel_characterizing_2024}%
  \BibitemOpen
  \bibfield  {author} {\bibinfo {author} {\bibfnamefont {S.}~\bibnamefont
  {Hiebel}}, \bibinfo {author} {\bibfnamefont {B.}~\bibnamefont {Nagler}},
  \bibinfo {author} {\bibfnamefont {S.}~\bibnamefont {Barbosa}}, \bibinfo
  {author} {\bibfnamefont {J.}~\bibnamefont {Koch}},\ and\ \bibinfo {author}
  {\bibfnamefont {A.}~\bibnamefont {Widera}},\ }\bibfield  {title} {\bibinfo
  {title} {Characterizing quantum gases in time-controlled disorder
  realizations using cross-correlations of density distributions},\ }\href
  {https://dx.doi.org/10.1088/1367-2630/ad1b82} {\bibfield  {journal} {\bibinfo
   {journal} {New Journal of Physics}\ }\textbf {\bibinfo {volume} {26}},\
  \bibinfo {pages} {013042} (\bibinfo {year} {2024})}\BibitemShut {NoStop}%
\bibitem [{\citenamefont {Gänger}\ \emph {et~al.}(2018)\citenamefont
  {Gänger}, \citenamefont {Phieler}, \citenamefont {Nagler},\ and\
  \citenamefont {Widera}}]{ganger_versatile_2018}%
  \BibitemOpen
  \bibfield  {author} {\bibinfo {author} {\bibfnamefont {B.}~\bibnamefont
  {Gänger}}, \bibinfo {author} {\bibfnamefont {J.}~\bibnamefont {Phieler}},
  \bibinfo {author} {\bibfnamefont {B.}~\bibnamefont {Nagler}},\ and\ \bibinfo
  {author} {\bibfnamefont {A.}~\bibnamefont {Widera}},\ }\bibfield  {title}
  {\bibinfo {title} {A versatile apparatus for fermionic lithium quantum gases
  based on an interference-filter laser system},\ }\href
  {https://doi.org/10.1063/1.5045827} {\bibfield  {journal} {\bibinfo
  {journal} {Review of Scientific Instruments}\ }\textbf {\bibinfo {volume}
  {89}},\ \bibinfo {pages} {093105} (\bibinfo {year} {2018})}\BibitemShut
  {NoStop}%
\bibitem [{\citenamefont {Nagler}\ \emph
  {et~al.}(2022{\natexlab{b}})\citenamefont {Nagler}, \citenamefont {Barbosa},
  \citenamefont {Koch}, \citenamefont {Orso},\ and\ \citenamefont
  {Widera}}]{nagler_observing_2022}%
  \BibitemOpen
  \bibfield  {author} {\bibinfo {author} {\bibfnamefont {B.}~\bibnamefont
  {Nagler}}, \bibinfo {author} {\bibfnamefont {S.}~\bibnamefont {Barbosa}},
  \bibinfo {author} {\bibfnamefont {J.}~\bibnamefont {Koch}}, \bibinfo {author}
  {\bibfnamefont {G.}~\bibnamefont {Orso}},\ and\ \bibinfo {author}
  {\bibfnamefont {A.}~\bibnamefont {Widera}},\ }\bibfield  {title} {\bibinfo
  {title} {Observing the loss and revival of long-range phase coherence through
  disorder quenches},\ }\href {https://www.pnas.org/content/119/1/e2111078118}
  {\bibfield  {journal} {\bibinfo  {journal} {Proceedings of the National
  Academy of Sciences}\ }\textbf {\bibinfo {volume} {119}} (\bibinfo {year}
  {2022}{\natexlab{b}})}\BibitemShut {NoStop}%
\bibitem [{\citenamefont {Kuhn}\ \emph {et~al.}(2007)\citenamefont {Kuhn},
  \citenamefont {Sigwarth}, \citenamefont {Miniatura}, \citenamefont
  {Delande},\ and\ \citenamefont {Müller}}]{kuhn_coherent_2007}%
  \BibitemOpen
  \bibfield  {author} {\bibinfo {author} {\bibfnamefont {R.~C.}\ \bibnamefont
  {Kuhn}}, \bibinfo {author} {\bibfnamefont {O.}~\bibnamefont {Sigwarth}},
  \bibinfo {author} {\bibfnamefont {C.}~\bibnamefont {Miniatura}}, \bibinfo
  {author} {\bibfnamefont {D.}~\bibnamefont {Delande}},\ and\ \bibinfo {author}
  {\bibfnamefont {C.~A.}\ \bibnamefont {Müller}},\ }\bibfield  {title}
  {\bibinfo {title} {Coherent matter wave transport in speckle potentials},\
  }\href {https://doi.org/10.1088/1367-2630/9/6/161} {\bibfield  {journal}
  {\bibinfo  {journal} {New Journal of Physics}\ }\textbf {\bibinfo {volume}
  {9}},\ \bibinfo {pages} {161} (\bibinfo {year} {2007})}\BibitemShut {NoStop}%
\bibitem [{\citenamefont {Pilati}\ \emph {et~al.}(2010)\citenamefont {Pilati},
  \citenamefont {Giorgini}, \citenamefont {Modugno},\ and\ \citenamefont
  {Prokof'ev}}]{pilati_dilute_2010}%
  \BibitemOpen
  \bibfield  {author} {\bibinfo {author} {\bibfnamefont {S.}~\bibnamefont
  {Pilati}}, \bibinfo {author} {\bibfnamefont {S.}~\bibnamefont {Giorgini}},
  \bibinfo {author} {\bibfnamefont {M.}~\bibnamefont {Modugno}},\ and\ \bibinfo
  {author} {\bibfnamefont {N.}~\bibnamefont {Prokof'ev}},\ }\bibfield  {title}
  {\bibinfo {title} {Dilute {Bose} gas with correlated disorder: a path
  integral {Monte} {Carlo} study},\ }\href
  {https://doi.org/10.1088/1367-2630/12/7/073003} {\bibfield  {journal}
  {\bibinfo  {journal} {New Journal of Physics}\ }\textbf {\bibinfo {volume}
  {12}},\ \bibinfo {pages} {073003} (\bibinfo {year} {2010})}\BibitemShut
  {NoStop}%
\bibitem [{\citenamefont {Sanchez-Palencia}\ \emph {et~al.}(2008)\citenamefont
  {Sanchez-Palencia}, \citenamefont {Clément}, \citenamefont {Lugan},
  \citenamefont {Bouyer},\ and\ \citenamefont
  {Aspect}}]{sanchez-palencia_disorder-induced_2008}%
  \BibitemOpen
  \bibfield  {author} {\bibinfo {author} {\bibfnamefont {L.}~\bibnamefont
  {Sanchez-Palencia}}, \bibinfo {author} {\bibfnamefont {D.}~\bibnamefont
  {Clément}}, \bibinfo {author} {\bibfnamefont {P.}~\bibnamefont {Lugan}},
  \bibinfo {author} {\bibfnamefont {P.}~\bibnamefont {Bouyer}},\ and\ \bibinfo
  {author} {\bibfnamefont {A.}~\bibnamefont {Aspect}},\ }\bibfield  {title}
  {\bibinfo {title} {Disorder-induced trapping versus {Anderson} localization
  in {Bose}–{Einstein} condensates expanding in disordered potentials},\
  }\href {https://doi.org/10.1088/1367-2630/10/4/045019} {\bibfield  {journal}
  {\bibinfo  {journal} {New Journal of Physics}\ }\textbf {\bibinfo {volume}
  {10}},\ \bibinfo {pages} {045019} (\bibinfo {year} {2008})}\BibitemShut
  {NoStop}%
\bibitem [{\citenamefont {Barbosa}\ \emph {et~al.}(2023)\citenamefont
  {Barbosa}, \citenamefont {Kiefer-Emmanouilidis}, \citenamefont {Lang},
  \citenamefont {Koch},\ and\ \citenamefont {Widera}}]{barbosa_what_2023}%
  \BibitemOpen
  \bibfield  {author} {\bibinfo {author} {\bibfnamefont {S.}~\bibnamefont
  {Barbosa}}, \bibinfo {author} {\bibfnamefont {M.}~\bibnamefont
  {Kiefer-Emmanouilidis}}, \bibinfo {author} {\bibfnamefont {F.}~\bibnamefont
  {Lang}}, \bibinfo {author} {\bibfnamefont {J.}~\bibnamefont {Koch}},\ and\
  \bibinfo {author} {\bibfnamefont {A.}~\bibnamefont {Widera}},\ }\bibfield
  {title} {\bibinfo {title} {What can we learn from diffusion about {Anderson}
  localization of a degenerate {Fermi} gas?},\ }\href
  {http://arxiv.org/abs/2311.07505} {\bibfield  {journal} {\bibinfo  {journal}
  {arXiv:2311.07505}\ } (\bibinfo {year} {2023})}\BibitemShut {NoStop}%
\bibitem [{\citenamefont {Beilin}\ \emph {et~al.}(2010)\citenamefont {Beilin},
  \citenamefont {Gurevich},\ and\ \citenamefont
  {Shapiro}}]{beilin_diffusion_2010}%
  \BibitemOpen
  \bibfield  {author} {\bibinfo {author} {\bibfnamefont {L.}~\bibnamefont
  {Beilin}}, \bibinfo {author} {\bibfnamefont {E.}~\bibnamefont {Gurevich}},\
  and\ \bibinfo {author} {\bibfnamefont {B.}~\bibnamefont {Shapiro}},\
  }\bibfield  {title} {\bibinfo {title} {Diffusion of cold-atomic gases in the
  presence of an optical speckle potential},\ }\href
  {https://doi.org/10.1103/PhysRevA.81.033612} {\bibfield  {journal} {\bibinfo
  {journal} {Physical Review A}\ }\textbf {\bibinfo {volume} {81}},\ \bibinfo
  {pages} {033612} (\bibinfo {year} {2010})}\BibitemShut {NoStop}%
\bibitem [{\citenamefont {Koch}\ \emph {et~al.}(2023)\citenamefont {Koch},
  \citenamefont {Menon}, \citenamefont {Cuestas}, \citenamefont {Barbosa},
  \citenamefont {Lutz}, \citenamefont {Fogarty}, \citenamefont {Busch},\ and\
  \citenamefont {Widera}}]{koch_quantum_2023}%
  \BibitemOpen
  \bibfield  {author} {\bibinfo {author} {\bibfnamefont {J.}~\bibnamefont
  {Koch}}, \bibinfo {author} {\bibfnamefont {K.}~\bibnamefont {Menon}},
  \bibinfo {author} {\bibfnamefont {E.}~\bibnamefont {Cuestas}}, \bibinfo
  {author} {\bibfnamefont {S.}~\bibnamefont {Barbosa}}, \bibinfo {author}
  {\bibfnamefont {E.}~\bibnamefont {Lutz}}, \bibinfo {author} {\bibfnamefont
  {T.}~\bibnamefont {Fogarty}}, \bibinfo {author} {\bibfnamefont
  {T.}~\bibnamefont {Busch}},\ and\ \bibinfo {author} {\bibfnamefont
  {A.}~\bibnamefont {Widera}},\ }\bibfield  {title} {\bibinfo {title} {A
  quantum engine in the {BEC}–{BCS} crossover},\ }\href
  {https://doi.org/10.1038/s41586-023-06469-8} {\bibfield  {journal} {\bibinfo
  {journal} {Nature}\ }\textbf {\bibinfo {volume} {621}},\ \bibinfo {pages}
  {723} (\bibinfo {year} {2023})}\BibitemShut {NoStop}%
\bibitem [{\citenamefont {Barbosa}\ \emph {et~al.}()\citenamefont {Barbosa},
  \citenamefont {Kiefer-Emmanouilidis}, \citenamefont {Lang}, \citenamefont
  {Koch},\ and\ \citenamefont {Widera}}]{zenodo_dynamic}%
  \BibitemOpen
  \bibfield  {author} {\bibinfo {author} {\bibfnamefont {S.}~\bibnamefont
  {Barbosa}}, \bibinfo {author} {\bibfnamefont {M.}~\bibnamefont
  {Kiefer-Emmanouilidis}}, \bibinfo {author} {\bibfnamefont {F.}~\bibnamefont
  {Lang}}, \bibinfo {author} {\bibfnamefont {J.}~\bibnamefont {Koch}},\ and\
  \bibinfo {author} {\bibfnamefont {A.}~\bibnamefont {Widera}},\ }\href
  {https://zenodo.org/doi/10.5281/zenodo.10478890} {\bibinfo {title} {Zenodo
  repository: https://zenodo.org/doi/10.5281/zenodo.10478890}}\BibitemShut
  {NoStop}%
\end{thebibliography}

\begin{thebibliography}{18}%
\makeatletter
\providecommand \@ifxundefined [1]{%
 \@ifx{#1\undefined}
}%
\providecommand \@ifnum [1]{%
 \ifnum #1\expandafter \@firstoftwo
 \else \expandafter \@secondoftwo
 \fi
}%
\providecommand \@ifx [1]{%
 \ifx #1\expandafter \@firstoftwo
 \else \expandafter \@secondoftwo
 \fi
}%
\providecommand \natexlab [1]{#1}%
\providecommand \enquote  [1]{``#1''}%
\providecommand \bibnamefont  [1]{#1}%
\providecommand \bibfnamefont [1]{#1}%
\providecommand \citenamefont [1]{#1}%
\providecommand \href@noop [0]{\@secondoftwo}%
\providecommand \href [0]{\begingroup \@sanitize@url \@href}%
\providecommand \@href[1]{\@@startlink{#1}\@@href}%
\providecommand \@@href[1]{\endgroup#1\@@endlink}%
\providecommand \@sanitize@url [0]{\catcode `\\12\catcode `\$12\catcode
  `\&12\catcode `\#12\catcode `\^12\catcode `\_12\catcode `\%12\relax}%
\providecommand \@@startlink[1]{}%
\providecommand \@@endlink[0]{}%
\providecommand \url  [0]{\begingroup\@sanitize@url \@url }%
\providecommand \@url [1]{\endgroup\@href {#1}{\urlprefix }}%
\providecommand \urlprefix  [0]{URL }%
\providecommand \Eprint [0]{\href }%
\providecommand \doibase [0]{https://doi.org/}%
\providecommand \selectlanguage [0]{\@gobble}%
\providecommand \bibinfo  [0]{\@secondoftwo}%
\providecommand \bibfield  [0]{\@secondoftwo}%
\providecommand \translation [1]{[#1]}%
\providecommand \BibitemOpen [0]{}%
\providecommand \bibitemStop [0]{}%
\providecommand \bibitemNoStop [0]{.\EOS\space}%
\providecommand \EOS [0]{\spacefactor3000\relax}%
\providecommand \BibitemShut  [1]{\csname bibitem#1\endcsname}%
\let\auto@bib@innerbib\@empty
%</preamble>
\bibitem [{\citenamefont {Barbosa}\ \emph {et~al.}(2023)\citenamefont
  {Barbosa}, \citenamefont {Kiefer-Emmanouilidis}, \citenamefont {Lang},
  \citenamefont {Koch},\ and\ \citenamefont {Widera}}]{barbosa_what_2023S}%
  \BibitemOpen
  \bibfield  {author} {\bibinfo {author} {\bibfnamefont {S.}~\bibnamefont
  {Barbosa}}, \bibinfo {author} {\bibfnamefont {M.}~\bibnamefont
  {Kiefer-Emmanouilidis}}, \bibinfo {author} {\bibfnamefont {F.}~\bibnamefont
  {Lang}}, \bibinfo {author} {\bibfnamefont {J.}~\bibnamefont {Koch}},\ and\
  \bibinfo {author} {\bibfnamefont {A.}~\bibnamefont {Widera}},\ }\bibfield
  {title} {\bibinfo {title} {What can we learn from diffusion about {Anderson}
  localization of a degenerate {Fermi} gas?},\ }\href
  {http://arxiv.org/abs/2311.07505} {\bibfield  {journal} {\bibinfo  {journal}
  {arXiv:2311.07505}\ } (\bibinfo {year} {2023})}\BibitemShut {NoStop}%
\bibitem [{\citenamefont {Reinaudi}\ \emph {et~al.}(2007)\citenamefont
  {Reinaudi}, \citenamefont {Lahaye}, \citenamefont {Wang},\ and\ \citenamefont
  {Guéry-Odelin}}]{reinaudi_strong_2007S}%
  \BibitemOpen
  \bibfield  {author} {\bibinfo {author} {\bibfnamefont {G.}~\bibnamefont
  {Reinaudi}}, \bibinfo {author} {\bibfnamefont {T.}~\bibnamefont {Lahaye}},
  \bibinfo {author} {\bibfnamefont {Z.}~\bibnamefont {Wang}},\ and\ \bibinfo
  {author} {\bibfnamefont {D.}~\bibnamefont {Guéry-Odelin}},\ }\bibfield
  {title} {\bibinfo {title} {Strong saturation absorption imaging of dense
  clouds of ultracold atoms},\ }\href {https://doi.org/10.1364/OL.32.003143}
  {\bibfield  {journal} {\bibinfo  {journal} {Optics Letters}\ }\textbf
  {\bibinfo {volume} {32}},\ \bibinfo {pages} {3143} (\bibinfo {year}
  {2007})}\BibitemShut {NoStop}%
\bibitem [{\citenamefont {Nagler}\ \emph {et~al.}(2020)\citenamefont {Nagler},
  \citenamefont {Radonjić}, \citenamefont {Barbosa}, \citenamefont {Koch},
  \citenamefont {Pelster},\ and\ \citenamefont {Widera}}]{nagler_cloud_2020S}%
  \BibitemOpen
  \bibfield  {author} {\bibinfo {author} {\bibfnamefont {B.}~\bibnamefont
  {Nagler}}, \bibinfo {author} {\bibfnamefont {M.}~\bibnamefont {Radonjić}},
  \bibinfo {author} {\bibfnamefont {S.}~\bibnamefont {Barbosa}}, \bibinfo
  {author} {\bibfnamefont {J.}~\bibnamefont {Koch}}, \bibinfo {author}
  {\bibfnamefont {A.}~\bibnamefont {Pelster}},\ and\ \bibinfo {author}
  {\bibfnamefont {A.}~\bibnamefont {Widera}},\ }\bibfield  {title} {\bibinfo
  {title} {Cloud shape of a molecular {Bose}–{Einstein} condensate in a
  disordered trap: a case study of the dirty boson problem},\ }\href
  {https://doi.org/10.1088/1367-2630/ab73cb} {\bibfield  {journal} {\bibinfo
  {journal} {New Journal of Physics}\ }\textbf {\bibinfo {volume} {22}},\
  \bibinfo {pages} {033021} (\bibinfo {year} {2020})}\BibitemShut {NoStop}%
\bibitem [{\citenamefont {Gänger}\ \emph {et~al.}(2018)\citenamefont
  {Gänger}, \citenamefont {Phieler}, \citenamefont {Nagler},\ and\
  \citenamefont {Widera}}]{ganger_versatile_2018S}%
  \BibitemOpen
  \bibfield  {author} {\bibinfo {author} {\bibfnamefont {B.}~\bibnamefont
  {Gänger}}, \bibinfo {author} {\bibfnamefont {J.}~\bibnamefont {Phieler}},
  \bibinfo {author} {\bibfnamefont {B.}~\bibnamefont {Nagler}},\ and\ \bibinfo
  {author} {\bibfnamefont {A.}~\bibnamefont {Widera}},\ }\bibfield  {title}
  {\bibinfo {title} {A versatile apparatus for fermionic lithium quantum gases
  based on an interference-filter laser system},\ }\href
  {https://doi.org/10.1063/1.5045827} {\bibfield  {journal} {\bibinfo
  {journal} {Review of Scientific Instruments}\ }\textbf {\bibinfo {volume}
  {89}},\ \bibinfo {pages} {093105} (\bibinfo {year} {2018})}\BibitemShut
  {NoStop}%
\bibitem [{\citenamefont {Grimm}(2007)}]{grimm_ultracold_2007S}%
  \BibitemOpen
  \bibfield  {author} {\bibinfo {author} {\bibfnamefont {R.}~\bibnamefont
  {Grimm}},\ }\bibfield  {title} {\bibinfo {title} {Ultracold {Fermi} gases in
  the {BEC}-{BCS} crossover: a review from the {Innsbruck} perspective},\
  }\href {http://arxiv.org/abs/cond-mat/0703091} {\bibfield  {journal}
  {\bibinfo  {journal} {arXiv:cond-mat/0703091}\ } (\bibinfo {year}
  {2007})}\BibitemShut {NoStop}%
\bibitem [{\citenamefont {Zürn}\ \emph {et~al.}(2013)\citenamefont {Zürn},
  \citenamefont {Lompe}, \citenamefont {Wenz}, \citenamefont {Jochim},
  \citenamefont {Julienne},\ and\ \citenamefont {Hutson}}]{zurn_precise_2013S}%
  \BibitemOpen
  \bibfield  {author} {\bibinfo {author} {\bibfnamefont {G.}~\bibnamefont
  {Zürn}}, \bibinfo {author} {\bibfnamefont {T.}~\bibnamefont {Lompe}},
  \bibinfo {author} {\bibfnamefont {A.~N.}\ \bibnamefont {Wenz}}, \bibinfo
  {author} {\bibfnamefont {S.}~\bibnamefont {Jochim}}, \bibinfo {author}
  {\bibfnamefont {P.~S.}\ \bibnamefont {Julienne}},\ and\ \bibinfo {author}
  {\bibfnamefont {J.~M.}\ \bibnamefont {Hutson}},\ }\bibfield  {title}
  {\bibinfo {title} {Precise {Characterization} of $^6${Li} {Feshbach}
  {Resonances} {Using} {Trap}-{Sideband}-{Resolved} {RF} {Spectroscopy} of
  {Weakly} {Bound} {Molecules}},\ }\href
  {https://doi.org/10.1103/PhysRevLett.110.135301} {\bibfield  {journal}
  {\bibinfo  {journal} {Physical Review Letters}\ }\textbf {\bibinfo {volume}
  {110}},\ \bibinfo {pages} {135301} (\bibinfo {year} {2013})}\BibitemShut
  {NoStop}%
\bibitem [{\citenamefont {Nagler}\ \emph
  {et~al.}(2022{\natexlab{a}})\citenamefont {Nagler}, \citenamefont {Barbosa},
  \citenamefont {Koch}, \citenamefont {Orso},\ and\ \citenamefont
  {Widera}}]{nagler_observing_2022S}%
  \BibitemOpen
  \bibfield  {author} {\bibinfo {author} {\bibfnamefont {B.}~\bibnamefont
  {Nagler}}, \bibinfo {author} {\bibfnamefont {S.}~\bibnamefont {Barbosa}},
  \bibinfo {author} {\bibfnamefont {J.}~\bibnamefont {Koch}}, \bibinfo {author}
  {\bibfnamefont {G.}~\bibnamefont {Orso}},\ and\ \bibinfo {author}
  {\bibfnamefont {A.}~\bibnamefont {Widera}},\ }\bibfield  {title} {\bibinfo
  {title} {Observing the loss and revival of long-range phase coherence through
  disorder quenches},\ }\href {https://www.pnas.org/content/119/1/e2111078118}
  {\bibfield  {journal} {\bibinfo  {journal} {Proceedings of the National
  Academy of Sciences}\ }\textbf {\bibinfo {volume} {119}} (\bibinfo {year}
  {2022}{\natexlab{a}})}\BibitemShut {NoStop}%
\bibitem [{\citenamefont {Hadzibabic}\ \emph {et~al.}(2002)\citenamefont
  {Hadzibabic}, \citenamefont {Stan}, \citenamefont {Dieckmann}, \citenamefont
  {Gupta}, \citenamefont {Zwierlein}, \citenamefont {G{\"o}rlitz},\ and\
  \citenamefont {Ketterle}}]{hadzibabic_twospecies_2002S}%
  \BibitemOpen
  \bibfield  {author} {\bibinfo {author} {\bibfnamefont {Z.}~\bibnamefont
  {Hadzibabic}}, \bibinfo {author} {\bibfnamefont {C.}~\bibnamefont {Stan}},
  \bibinfo {author} {\bibfnamefont {K.}~\bibnamefont {Dieckmann}}, \bibinfo
  {author} {\bibfnamefont {S.}~\bibnamefont {Gupta}}, \bibinfo {author}
  {\bibfnamefont {M.}~\bibnamefont {Zwierlein}}, \bibinfo {author}
  {\bibfnamefont {A.}~\bibnamefont {G{\"o}rlitz}},\ and\ \bibinfo {author}
  {\bibfnamefont {W.}~\bibnamefont {Ketterle}},\ }\bibfield  {title} {\bibinfo
  {title} {Two-species mixture of quantum degenerate bose and fermi gases},\
  }\href {https://journals.aps.org/prl/abstract/10.1103/PhysRevLett.88.160401}
  {\bibfield  {journal} {\bibinfo  {journal} {Physical Review Letters}\
  }\textbf {\bibinfo {volume} {88}},\ \bibinfo {pages} {160401} (\bibinfo
  {year} {2002})}\BibitemShut {NoStop}%
\bibitem [{\citenamefont {Kinast}(2006)}]{kinast_phdS}%
  \BibitemOpen
  \bibfield  {author} {\bibinfo {author} {\bibfnamefont {J.~M.}\ \bibnamefont
  {Kinast}},\ }\emph {\bibinfo {title} {Thermodynamics and superfluidity of a
  strongly interacting Fermi gas}},\ \href@noop {} {Ph.D. thesis},\ \bibinfo
  {school} {Duke University} (\bibinfo {year} {2006})\BibitemShut {NoStop}%
\bibitem [{\citenamefont {Bouchet}\ \emph {et~al.}(2004)\citenamefont
  {Bouchet}, \citenamefont {Cecconi},\ and\ \citenamefont
  {Vulpiani}}]{bouchet_minimal_2004S}%
  \BibitemOpen
  \bibfield  {author} {\bibinfo {author} {\bibfnamefont {F.}~\bibnamefont
  {Bouchet}}, \bibinfo {author} {\bibfnamefont {F.}~\bibnamefont {Cecconi}},\
  and\ \bibinfo {author} {\bibfnamefont {A.}~\bibnamefont {Vulpiani}},\
  }\bibfield  {title} {\bibinfo {title} {Minimal {Stochastic} {Model} for
  {Fermi}'s {Acceleration}},\ }\href
  {https://doi.org/10.1103/PhysRevLett.92.040601} {\bibfield  {journal}
  {\bibinfo  {journal} {Physical Review Letters}\ }\textbf {\bibinfo {volume}
  {92}},\ \bibinfo {pages} {040601} (\bibinfo {year} {2004})}\BibitemShut
  {NoStop}%
\bibitem [{\citenamefont {Golubović}\ \emph {et~al.}(1991)\citenamefont
  {Golubović}, \citenamefont {Feng},\ and\ \citenamefont
  {Zeng}}]{golubovic_classical_1991S}%
  \BibitemOpen
  \bibfield  {author} {\bibinfo {author} {\bibfnamefont {L.}~\bibnamefont
  {Golubović}}, \bibinfo {author} {\bibfnamefont {S.}~\bibnamefont {Feng}},\
  and\ \bibinfo {author} {\bibfnamefont {F.-A.}\ \bibnamefont {Zeng}},\
  }\bibfield  {title} {\bibinfo {title} {Classical and quantum superdiffusion
  in a time-dependent random potential},\ }\href
  {https://doi.org/10.1103/PhysRevLett.67.2115} {\bibfield  {journal} {\bibinfo
   {journal} {Physical Review Letters}\ }\textbf {\bibinfo {volume} {67}},\
  \bibinfo {pages} {2115} (\bibinfo {year} {1991})}\BibitemShut {NoStop}%
\bibitem [{\citenamefont {Rosenbluth}(1992)}]{rosenbluth_comment_1992S}%
  \BibitemOpen
  \bibfield  {author} {\bibinfo {author} {\bibfnamefont {M.~N.}\ \bibnamefont
  {Rosenbluth}},\ }\bibfield  {title} {\bibinfo {title} {Comment on
  ``{Classical} and quantum superdiffusion in a time-dependent random
  potential''},\ }\href {https://doi.org/10.1103/PhysRevLett.69.1831}
  {\bibfield  {journal} {\bibinfo  {journal} {Physical Review Letters}\
  }\textbf {\bibinfo {volume} {69}},\ \bibinfo {pages} {1831} (\bibinfo {year}
  {1992})}\BibitemShut {NoStop}%
\bibitem [{\citenamefont {Aguer}\ \emph {et~al.}(2010)\citenamefont {Aguer},
  \citenamefont {De~Bièvre}, \citenamefont {Lafitte},\ and\ \citenamefont
  {Parris}}]{aguer_classical_2010S}%
  \BibitemOpen
  \bibfield  {author} {\bibinfo {author} {\bibfnamefont {B.}~\bibnamefont
  {Aguer}}, \bibinfo {author} {\bibfnamefont {S.}~\bibnamefont {De~Bièvre}},
  \bibinfo {author} {\bibfnamefont {P.}~\bibnamefont {Lafitte}},\ and\ \bibinfo
  {author} {\bibfnamefont {P.~E.}\ \bibnamefont {Parris}},\ }\bibfield  {title}
  {\bibinfo {title} {Classical {Motion} in {Force} {Fields} with {Short}
  {Range} {Correlations}},\ }\href {https://doi.org/10.1007/s10955-009-9898-7}
  {\bibfield  {journal} {\bibinfo  {journal} {Journal of Statistical Physics}\
  }\textbf {\bibinfo {volume} {138}},\ \bibinfo {pages} {780} (\bibinfo {year}
  {2010})}\BibitemShut {NoStop}%
\bibitem [{\citenamefont {Levi}\ \emph {et~al.}(2012)\citenamefont {Levi},
  \citenamefont {Krivolapov}, \citenamefont {Fishman},\ and\ \citenamefont
  {Segev}}]{levi_hyper-transport_2012S}%
  \BibitemOpen
  \bibfield  {author} {\bibinfo {author} {\bibfnamefont {L.}~\bibnamefont
  {Levi}}, \bibinfo {author} {\bibfnamefont {Y.}~\bibnamefont {Krivolapov}},
  \bibinfo {author} {\bibfnamefont {S.}~\bibnamefont {Fishman}},\ and\ \bibinfo
  {author} {\bibfnamefont {M.}~\bibnamefont {Segev}},\ }\bibfield  {title}
  {\bibinfo {title} {Hyper-transport of light and stochastic acceleration by
  evolving disorder},\ }\href {https://doi.org/10.1038/nphys2463} {\bibfield
  {journal} {\bibinfo  {journal} {Nature Physics}\ }\textbf {\bibinfo {volume}
  {8}},\ \bibinfo {pages} {912} (\bibinfo {year} {2012})}\BibitemShut {NoStop}%
\bibitem [{\citenamefont {Nagler}\ \emph
  {et~al.}(2022{\natexlab{b}})\citenamefont {Nagler}, \citenamefont {Will},
  \citenamefont {Hiebel}, \citenamefont {Barbosa}, \citenamefont {Koch},
  \citenamefont {Fleischhauer},\ and\ \citenamefont
  {Widera}}]{nagler_ultracold_2022S}%
  \BibitemOpen
  \bibfield  {author} {\bibinfo {author} {\bibfnamefont {B.}~\bibnamefont
  {Nagler}}, \bibinfo {author} {\bibfnamefont {M.}~\bibnamefont {Will}},
  \bibinfo {author} {\bibfnamefont {S.}~\bibnamefont {Hiebel}}, \bibinfo
  {author} {\bibfnamefont {S.}~\bibnamefont {Barbosa}}, \bibinfo {author}
  {\bibfnamefont {J.}~\bibnamefont {Koch}}, \bibinfo {author} {\bibfnamefont
  {M.}~\bibnamefont {Fleischhauer}},\ and\ \bibinfo {author} {\bibfnamefont
  {A.}~\bibnamefont {Widera}},\ }\bibfield  {title} {\bibinfo {title}
  {Ultracold {Bose} {Gases} in {Dynamic} {Disorder} with {Tunable}
  {Correlation} {Time}},\ }\href
  {https://doi.org/10.1103/PhysRevLett.128.233601} {\bibfield  {journal}
  {\bibinfo  {journal} {Physical Review Letters}\ }\textbf {\bibinfo {volume}
  {128}},\ \bibinfo {pages} {233601} (\bibinfo {year}
  {2022}{\natexlab{b}})}\BibitemShut {NoStop}%
\bibitem [{\citenamefont {Volpe}\ \emph {et~al.}(2014)\citenamefont {Volpe},
  \citenamefont {Volpe},\ and\ \citenamefont {Gigan}}]{volpe_brownian_2014S}%
  \BibitemOpen
  \bibfield  {author} {\bibinfo {author} {\bibfnamefont {G.}~\bibnamefont
  {Volpe}}, \bibinfo {author} {\bibfnamefont {G.}~\bibnamefont {Volpe}},\ and\
  \bibinfo {author} {\bibfnamefont {S.}~\bibnamefont {Gigan}},\ }\bibfield
  {title} {\bibinfo {title} {Brownian {Motion} in a {Speckle} {Light} {Field}:
  {Tunable} {Anomalous} {Diffusion} and {Selective} {Optical} {Manipulation}},\
  }\href {https://doi.org/10.1038/srep03936} {\bibfield  {journal} {\bibinfo
  {journal} {Scientific Reports}\ }\textbf {\bibinfo {volume} {4}},\ \bibinfo
  {pages} {3936} (\bibinfo {year} {2014})}\BibitemShut {NoStop}%
\bibitem [{\citenamefont {Ostrowski}\ and\ \citenamefont
  {Siemieniec-Ozi\c{e}bło}(1997)}]{ostrowski_diffusion_1997S}%
  \BibitemOpen
  \bibfield  {author} {\bibinfo {author} {\bibfnamefont {M.}~\bibnamefont
  {Ostrowski}}\ and\ \bibinfo {author} {\bibfnamefont {G.}~\bibnamefont
  {Siemieniec-Ozi\c{e}bło}},\ }\bibfield  {title} {\bibinfo {title} {Diffusion
  in momentum space as a picture of second-order {Fermi} acceleration},\ }\href
  {https://doi.org/10.1016/S0927-6505(96)00061-8} {\bibfield  {journal}
  {\bibinfo  {journal} {Astroparticle Physics}\ }\textbf {\bibinfo {volume}
  {6}},\ \bibinfo {pages} {271} (\bibinfo {year} {1997})}\BibitemShut {NoStop}%
\bibitem [{\citenamefont {Jendrzejewski}\ \emph {et~al.}(2012)\citenamefont
  {Jendrzejewski}, \citenamefont {Bernard}, \citenamefont {Müller},
  \citenamefont {Cheinet}, \citenamefont {Josse}, \citenamefont {Piraud},
  \citenamefont {Pezzé}, \citenamefont {Sanchez-Palencia}, \citenamefont
  {Aspect},\ and\ \citenamefont
  {Bouyer}}]{jendrzejewski_three-dimensional_2012S}%
  \BibitemOpen
  \bibfield  {author} {\bibinfo {author} {\bibfnamefont {F.}~\bibnamefont
  {Jendrzejewski}}, \bibinfo {author} {\bibfnamefont {A.}~\bibnamefont
  {Bernard}}, \bibinfo {author} {\bibfnamefont {K.}~\bibnamefont {Müller}},
  \bibinfo {author} {\bibfnamefont {P.}~\bibnamefont {Cheinet}}, \bibinfo
  {author} {\bibfnamefont {V.}~\bibnamefont {Josse}}, \bibinfo {author}
  {\bibfnamefont {M.}~\bibnamefont {Piraud}}, \bibinfo {author} {\bibfnamefont
  {L.}~\bibnamefont {Pezzé}}, \bibinfo {author} {\bibfnamefont
  {L.}~\bibnamefont {Sanchez-Palencia}}, \bibinfo {author} {\bibfnamefont
  {A.}~\bibnamefont {Aspect}},\ and\ \bibinfo {author} {\bibfnamefont
  {P.}~\bibnamefont {Bouyer}},\ }\bibfield  {title} {\bibinfo {title}
  {Three-dimensional localization of ultracold atoms in an optical disordered
  potential},\ }\href {https://doi.org/10.1038/nphys2256} {\bibfield  {journal}
  {\bibinfo  {journal} {Nature Physics}\ }\textbf {\bibinfo {volume} {8}},\
  \bibinfo {pages} {398} (\bibinfo {year} {2012})}\BibitemShut {NoStop}%
\end{thebibliography}
\end{document}